\documentclass[journal]{IEEEtran}
\IEEEoverridecommandlockouts
\usepackage{cite}
\usepackage{amsmath,amssymb,amsfonts}
\usepackage{bbm}
\usepackage{algorithmic}
\usepackage{graphicx}
\usepackage{tabularx}
\usepackage{booktabs}
\usepackage{textcomp}
\usepackage{amsthm}
\usepackage{support-caption}
\usepackage{subcaption}
\usepackage{xcolor}
\def\BibTeX{{\rm B\kern-.05em{\sc i\kern-.025em b}\kern-.08em
    T\kern-.1667em\lower.7ex\hbox{E}\kern-.125emX}}

\usepackage{mathtools}
\usepackage{tikz}
\usetikzlibrary{shapes.arrows}
\usetikzlibrary{plotmarks}
\usetikzlibrary{patterns}
\usetikzlibrary{arrows.meta}
\usetikzlibrary{positioning}

\usepackage{pgfplots}
\pgfplotsset{compat=1.15}
\usepackage{adjustbox}

\usepackage{gincltex}
\usepgfplotslibrary{fillbetween}
\usepgfplotslibrary{groupplots}
\usetikzlibrary{plotmarks}
\usetikzlibrary{patterns}
\pgfplotsset{
tick label style={font=\footnotesize},
label style={font=\footnotesize},
legend style={font=\footnotesize},
}

\usetikzlibrary{external}
\definecolor{violet}{rgb}{0.6,0,0.6}%
\definecolor{orange_D}{rgb}{1,0.3,0}%
\definecolor{cyan}{rgb}{0,0.67,0.64}%
\definecolor{red}{rgb}{0.9,0,0}%
\definecolor{green}{rgb}{0,0.8,0}%
\definecolor{yellow}{rgb}{1,0.8,0}

\def \fwidth{0.8\linewidth}
\def \fheight {0.5\linewidth}

\DeclareMathOperator*{\argmin}{arg\,min}

\usepackage{glossaries}    
    
\newacronym[plural=MDPs,firstplural=Markov Decision Processes (MDPs)]{mdp}{MDP}{Markov Decision Process}
\newacronym{iot}{IoT}{Internet of Things}
\newacronym{fec}{FEC}{Forward Error Correction}
\newacronym{snr}{SNR}{Signal to Noise Ratio}
\newacronym{pmf}{pmf}{probability mass function}
\newacronym{iid}{IID}{Independent and Identically Distributed}
\newacronym{aoi}{AoI}{Age of Information}
\newacronym{eaoi}{EAoI}{Effective Age of Information}
\newacronym{paoi}{PAoI}{Peak Age of Information}
\newacronym{qaoi}{QAoI}{Age of Information at Query}
\newacronym{leo}{LEO}{Low Earth Orbit}
\newacronym{geo}{GEO}{Geosynchronous Earth Orbit}
\newacronym{pec}{PEC}{Packet Erasure Channel}
\newacronym{cdf}{CDF}{Cumulative Distribution Function}
\newacronym{pq}{PQ}{Permanent Query}
\newacronym{qapa}{QAPA}{Query Arrival Process Aware}
\newacronym{harq}{HARQ}{Hybrid Automated Repeat Request}
    
\newtheorem{theorem}{Theorem}
\newtheorem{lemma}{Lemma}

\renewcommand{\epsilon}{\varepsilon}
\begin{document}

\title{Query Age of Information: Freshness in\\ Pull-Based Communication}

\author{Federico Chiariotti, Josefine Holm, Anders E. Kal{\o}r, Beatriz Soret,\\ S{\o}ren K. Jensen, Torben B. Pedersen, and Petar Popovski\thanks{F. Chiariotti, J. Holm, A. E. Kal{\o}r, B. Soret, and P. Popovski are with the Department of Electronic Systems, Aalborg University (email: \{fchi,jho,aek,bsa,petarp\}@es.aau.dk). S K. Jensen and T. B. Pedersen are with the Department of Computer Science, Aalborg University (email: \{skj,tbp\}@cs.aau.dk). This work has been partly supported the Danish Council for Independent Research (Grant No. 8022-00284B SEMIOTIC).
}
}

\maketitle

\begin{abstract}
\gls{aoi} has become an important concept in communications, as it allows system designers to measure the freshness of the information available to remote monitoring or control processes. However, its definition tacitly assumes that new information is used at any time, which is not always the case: the instants at which information is collected and used may be dependent on a certain query process, and resource-constrained environments such as most \gls{iot} use cases require precise timing to fully exploit the limited available transmissions. In this work, we consider a \emph{pull-based communication model} in which the freshness of information is only important when the receiver generates a query: if the monitoring process is not using the value, the age of the last update is irrelevant. We optimize the \gls{qaoi}, a metric that samples the \gls{aoi} at relevant instants, better fitting the pull-based resource-constrained scenario, and show how this can lead to very different choices. Our results show that \gls{qaoi}-aware optimization can significantly reduce the average and worst-case perceived age for both periodic and stochastic queries.
\end{abstract}

\glsresetall

\begin{IEEEkeywords}
Age of Information, networked control systems, pull-based communication, Markov Decision Processes
\end{IEEEkeywords}

\section{Introduction}

Over the past few years, the concept of information freshness has received a significant attention in relation to cyber-physical systems that rely on communication of various updates in real time. This has led to the introduction of \emph{\gls{aoi}}~\cite{kosta2017age} as a measure that reflects the freshness at the receiver, and denotes the difference between the current time and the time when the most recently received update was generated at the sender.

In a common model for \gls{aoi}-sensitive systems, a wireless-equipped sensor measures a physical process and transmits its readings using a wireless link to a destination, where a monitor processes the received information. On the other hand, we study the effect of the existence of a \emph{query process} at the receiver in a resource-limited scenario in which a single sensor is constrained in how often it can transmit, either due to energy considerations or duty cycle limitations. In most works in the literature, the tacit assumption behind \gls{aoi} has been that the monitor at the receiver is interested in having fresh information at any time. In other words, the model assumes a \emph{push-based communication}, in which a hypothetical application residing at the monitor has a \emph{permanent query} to the updates that arrive at the receiver. Our work considers a \emph{pull-based communication} model, in which the query arrival process can guide the communication strategy by, for example, reading the sensor and transmitting the updates immediately before the query times. To design for pull-based communications, we introduce an age metric named \gls{qaoi}, which represents the age of the information available to the receiver in the instant when it actually needs it. This new metric is similar to the \gls{eaoi}, presented in~\cite{yin2019only}, which studied the effect of queries in a different scenario, with multi-user scheduling and a constant channel.
The \gls{qaoi} metric and the system optimization based on it is tested in a scenario with \emph{limited link availability} at the source, as our system model considers both when it is possible to transmit and when it is convenient to do so.

\begin{figure}[t]
    \centering
    \resizebox{\linewidth}{!}{\input{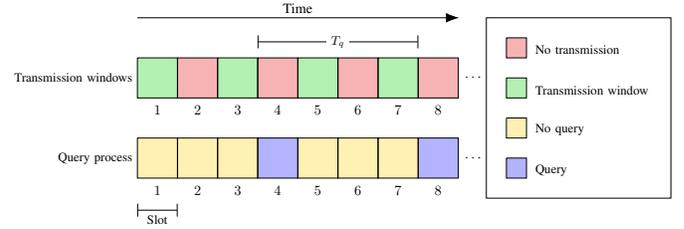}}
    %\tikzexternalenable
    \caption{Diagram of a satellite IoT-based pull-based communication model.}\vspace{-0.5cm}
    \label{fig:windows}
\end{figure}

One can think of several scenarios that would fit this pull-based model, as most monitoring and control applications operate over discrete time intervals, or only activate to react to some external trigger or user input. An interesting practical use case for this concept is represented by the Satellite \gls{iot}, which connects sensors in remote areas to the wider Internet through the use of \gls{geo} or \gls{leo} satellites, and is still a mostly unexplored setting for the \gls{aoi} literature~\cite{soret2020latency,li2019age}. Fig.~\ref{fig:windows} shows an example of the transmission and query timing with periodic intervals: the transmitter can only send a packet in the green slots, and the application queries the received result in the blue slots. In the push-based communication model, the transmitter should optimize for freshness at any time, while in the pull-based model, if there is a successful transmission in, e.g., slot 7, any transmission in slot 5 will then be useless to the application, as it will never see it. The transmitter will try to send packets as close to the query instant as possible, even if this results in a larger age in between queries.

\begin{figure*}[t!]
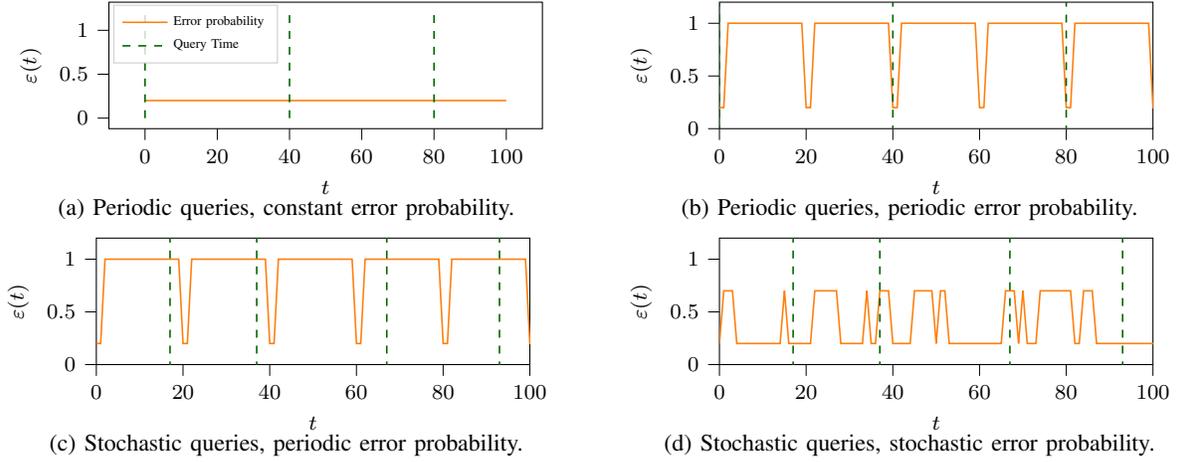

	\centering
	\begin{subfigure}[b]{0.45\linewidth}
	    \centering
        \input{JournalFigures/ErrorPlot1}
        \vspace{-0.3cm}\caption{Periodic queries, constant error probability.}\vspace{0.2cm}
        \label{fig:sc1}
    \end{subfigure}		
    \begin{subfigure}[b]{0.45\linewidth}
	    \centering
        \input{JournalFigures/ErrorPlot2}      
        \vspace{-0.3cm}\caption{Periodic queries, periodic error probability.}\vspace{0.2cm}
        \label{fig:sc2}
    \end{subfigure}	
    \begin{subfigure}[b]{0.45\linewidth}
	    \centering
        \input{JournalFigures/ErrorPlot3}       
        \vspace{-0.3cm}\caption{Stochastic queries, periodic error probability.}
        \label{fig:sc3}
    \end{subfigure}
        \begin{subfigure}[b]{0.45\linewidth}
	    \centering
        \input{RevisionFigures/ErrorPlot4}       
        \vspace{-0.3cm}\caption{Stochastic queries, stochastic error probability.}
        \label{fig:sc4}
    \end{subfigure}
 \caption{Scenarios considered in the simulations.}\vspace{-0.5cm}
 \label{fig:systems}
\end{figure*}

In our model, the channel between the sensor and the intermediate cache, i.e., the satellite uplink and downlink in the Satellite \gls{iot} scenario, is abstracted as a \gls{pec}, whose erasure probability can be either constant or time-varying. The query arrival process is, in general, a stochastic process. In our simulations, we consider four scenarios, which can correspond to three different Satellite \gls{iot} use cases:
\begin{itemize}
\item\emph{Periodic queries and constant channel}: this is the simplest case, in which queries are deterministic and periodic and the channel error probability is constant over time. This model can represent a \gls{geo} monitoring application, in which the satellite is always in the same position relative to the ground and the remote monitoring application simply logs the data by querying the ground station at predetermined intervals. This corresponds to the scenario in Fig.~\ref{fig:sc1}, in which $\epsilon(t)$ represents the channel error probability at time $t$. 
\item\emph{Periodic queries with a predictable channel}: in this case, we introduce some complexity in the channel behavior by having a time-varying error probability, while maintaining a deterministic and periodic query arrival process. This scenario can represent a \gls{leo} remote monitoring application, in which the sensor is not
served by a \gls{geo} satellite, but by a constellation of \gls{leo} satellites, whose orbits bring them outside the coverage range of the sensor: when there are no visible satellites, the packet error probability is 1. The sensors must then transmit its data when at least one satellite is passing over it. The orbits of the satellites can be computed in advance, so these periods of availability are known, but the sensor will be constrained in its scheduling decisions. This corresponds to the scenario in Fig.~\ref{fig:sc2}. Another possible example of this model would be a wireless scenario with recurring outages, during which the error probability for updates sharply increases.
\item\emph{Stochastic queries with a predictable channel}: in this case, the channel error probability can change over time and queries arrive according to a stochastic process, so the sensor will need to optimize the expected \gls{qaoi}, considering the probability of queries arriving in the near future. This scenario can represent a human-in-the-loop \gls{leo} monitoring application, in which queries are driven by the behavior of the user, and are then only partially predictable. This corresponds to the scenario in Fig.~\ref{fig:sc3}.
\item\emph{Stochastic queries with a stochastic channel}: this is the most general case, in which queries are stochastic, and the channel error probability is not predictable, but follows a stochastic process such as a Markov chain. This case can represent a general wireless channel, which does not have pre-computed satellite passes, but depends on the propagation conditions, and is shown in Fig.~\ref{fig:sc4}. It corresponds to a general wireless edge system getting queries from the Cloud.

\end{itemize}
These four examples are described in more detail and adapted to the communication model in Sec.~\ref{sec:results}. Besides the satellite IoT application of Fig.~\ref{fig:windows}, our model can fit several other monitoring applications, and the formulation is fully general for relaying scenarios with an intermediate cache node fulfilling requests from the end user. A generic example includes queries that are periodically/regularly sent from a central cloud to an edge node.

In this work, we model a scheduling problem for a resource-constrained sensor as a \gls{mdp}, showing the difference between a query-aware scheduler and a legacy one that tries to minimize \gls{aoi} at any instant: in the most general case, we consider the query arrival process and the channel state to be driven by two independent Markov chains. The model in this paper extends the framework we presented in~\cite{holm2020freshness}, which only considered a simple scenario with a constant channel and periodic queries. While previous works on \gls{aoi} often dealt with limitations on the link availability, because of congestion, energy, or propagation constraints, but to the best of our knowledge, this is the first work to combine limitations on the channel availability with \gls{aoi} in pull-based communication.

Our simulation results show that awareness of the query process can give significant gains in terms of both the average \gls{qaoi} and its higher percentiles, which represent a worst-case result and are critical for reliability-oriented applications, improving the performance of monitoring systems even in the most general case. The query-aware optimization often incurs a cost in terms of \gls{aoi}, as the scheduling strategy that optimizes freshness when a query arrives will often let the age increase when the probability of a query arriving is low,

The rest of this paper is divided as follows: first, Sec.~\ref{sec:related} presents the state of the art on scheduling and \gls{aoi} minimization. We then present the \gls{qaoi} metric and our communication system model in Sec.~\ref{sec:qaoi}, formulating it as an \gls{mdp} and finding the optimal policies in Sec.~\ref{sec:mdp}. We then present our simulation and its results in Sec.~\ref{sec:results}, considering a simple system first and gradually increasing its complexity. Finally, we conclude the paper and describe some possible avenues of future work in Sec.~\ref{sec:concl}.

\section{Related work}\label{sec:related}

Over the last few years, \gls{aoi}~\cite{kaul2012real} has attracted a significant amount of interest from the research community, as it represents a more relevant metric than latency for the ongoing monitoring and control of processes over a network. Most works in the literature deal with \gls{aoi} in queuing systems, examining different scheduling policies. Some recent works have proven that preemption, i.e., removing packets already in the queue in favor of newer ones with more up to date information, can give significant advantages in terms of average age~\cite{bedewy2017age}, even when multiple $M/M/1$ queuing systems in tandem are involved~\cite{bedewy2019multihop}. However, preemption can be suboptimal for different service time distributions, as the decision over whether to preempt or to continue the ongoing transmission becomes more complex~\cite{wang2018skip}.

Other works addressed \gls{aoi} in specific wireless scenarios with errors~\cite{chen2016error} and retransmissions~\cite{devassy2019reliable}, or basing their analysis on live experiments~\cite{beytur2019measuring}. The addition of more sources in the queuing system leads to an interesting scheduling problem, which aims at finding the packet generation rate that minimizes the age for the whole system~\cite{kadota2019minimizing}. Optimizing the access method and senders' updating policies to minimize \gls{aoi} in complex wireless communication system has been proven to be an NP-hard problem, but heuristics can achieve near-optimal solutions~\cite{sun2017update} by having sources decide whether an update is valuable enough to be sent (i.e., whether it would significantly reduce the \gls{aoi}) \cite{yates2015lazy}. The average \gls{aoi} has been derived in slotted~\cite{yates2017unreliable} and unslotted ALOHA~\cite{yates2020unslotted}, as well as in scheduled access~\cite{talak2018distributed}, and the performance of scheduling policies has been combined with these access methods in~\cite{chen2020rach}.

The scheduling problem can be formulated both for multiple sources, in which case the scheduling problem involves balancing the ages of the different sources while avoiding interference~\cite{talak2019optimizing}, or for a single source with resource constraints: usually, these constraints are in the form of limited energy availability or enforced duty cycles. Energy harvesting nodes are considered in~\cite{bacinoglu2018achieving}, which derives a near-optimal threshold-based heuristic that can work without knowledge of future energy generation, and by~\cite{wu2017optimal}, which derives the optimal policy for nodes with infinite as well as finite batteries. The trade-off between energy and freshness is examined explicitly in~\cite{gu2019timely}, while~\cite{feng2018optimal} considers a noisy channel as well, in which updates can be randomly erased. A more complex scenario, which includes a \gls{harq} channel as well as an energy harvesting node with a finite battery, is considered in~\cite{ceran2019reinforcement}, which models the problem as an \gls{mdp} and finds the optimal policy with reinforcement learning. Another interesting case for the scheduling problem is \gls{aoi} minimization in drone networks, in which drones have to move back and forth between the sensing location and the base station~\cite{zhang2020age}: finding the correct balance to minimize \gls{aoi} and energy expenditure is a non-trivial problem in this scenario.

\begin{table*}[t]\centering
    \footnotesize
	\begin{tabular}[c]{cl|cl}
		\toprule
		Symbol & Description & Symbol & Description \\
		\midrule
		$t_{u,i}$ & Delivery time of the $i$-th update & $T_q$ & Query arrival period \\
        $\Delta(t)$ & \gls{aoi} at time $t$ & $\mathcal{S}$ & State space of the scheduling \gls{mdp}\\
        $\Delta_{\infty}$ & Long-term expected \gls{aoi} & $\mathcal{A}$ & Action space of the scheduling \gls{mdp}\\
        $t_{q,i}$ & Time of the $i$-th query & $p_a(s,s')$ & Probability to go from $s$ to $s'$ for action $a$\\
        $\mathcal{S}_q$ & State space of the query process & $r(s,a,s')$ & Instantaneous reward\\
        $\mathcal{Q}$ & Set of states in which a query arrives & $b(t)$ & Number of available tokens at time $t$\\
        $P_q$ & Transition matrix of the query process & $c(s_t,a_t)$ & Long-term expected cost\\
        $\mathcal{S}_e$ & State space of the error probability process & $\lambda$ & Cost discount factor\\
        $P_e$ & Transition matrix of the error probability process & $\pi$ & Action policy\\ 
        $\epsilon(s_e)$ & Packet error probability for state $s_e$ & $v_{\pi}(s_t)$ & Expected state value with policy $\pi$\\
        $T_e$ & Packet error probability period & $\epsilon_0$ & Error probability during the satellite pass\\
		\bottomrule
	\end{tabular}
	\caption{Notation definitions.}
	\label{tab:notation}
\end{table*}

To the best of our knowledge, most of the literature so far has adopted a push-based model, in which updates are always relevant to the monitoring process.  We are aware of only a few other works that consider a pull-based model: the one most similar to this work~\cite{yin2019only} considers a server updating multiple users, using a metric called \gls{eaoi}, which is similar to \gls{qaoi}, although
in their case the system is not constrained by energy considerations, but by the presence of multiple sensors that need to be scheduled appropriately. The effect of queries is also modeled slightly differently: in our case, there is a strict ordering between transmissions and queries, and the response to a query is always sent immediately. On the other hand, \gls{eaoi} considers the possibility of a \emph{delayed response} if a transmission from the sensor is scheduled but not yet received. In our work, we focus on the optimization of the connection between the sensor and the intermediate cache, considering significant communication constraints and a more challenging \gls{iot} scenario. The difference between our metric and \gls{eaoi} is shown in Sec.~\ref{ssec:threshold}. Another work, first presented by Li \emph{et al.} in~\cite{sang2017power} and later expanded in~\cite{li2020waiting}, considers age not to matter unless and until a request for the information is generated. However, Li \emph{et al.}'s work does not consider the effects of scheduling in a regime with limited transmission opportunities, but rather tries to provide freshness to the user by combining multiple replications of the sensor value over multiple servers. The innovation from~\cite{li2020waiting} can be combined with ours with limited adaptations to the two models, resulting in a joint optimization of the whole end-to-end scheduling. Another work also models requests in the optimization function~\cite{hatami2020age}, but it only deals with memoryless request processes, which (as we will describe in the introduction) lead to a solution that is equivalent to standard \gls{aoi} minimization. The extended version of that paper~\cite{hatami2021aoi} considers more complex scenarios with partial battery knowledge, but still uses the same memoryless request model. Finally, a recent work by Xu \emph{et al.}~\cite{xu2020aoi} also considers a memoryless request process, but considers a mix between traditional \gls{aoi} and query-aware metrics. This is a significant difference from \gls{qaoi} and related metrics such as \gls{eaoi}, and it leads to a highly different optimization, which will prioritize users with a less active request process.

\section{System model}\label{sec:qaoi}
We now define a simple system model and consider the \gls{qaoi} metric. The notation in the following sections is summarized in Table~\ref{tab:notation}.

We consider a time-slotted system indexed by $t=1,2,\ldots$, and denote the time instants at which updates are successfully delivered to the edge node as $t_{u,1},t_{u,2},\ldots$. The source can be sampled at any time, and fresh information is always available, a condition known as zero-wait sampling. Following the common definition of \gls{aoi} considered in the literature, e.g. \cite{kaul2012real, kadota2019minimizing} we denote the \gls{aoi} in time slot $t$ by $\Delta(t)$, and define it as the difference between $t$ and the time at which the last successfully received packet was generated:
\begin{equation}
    \Delta(t)=t-\max_{i:t_{u,i}\le t} t_{u,i}.
\end{equation}
We will assume that $t_{u,1}=0$ so that $\Delta(t)$ is well defined. An alternative, but equivalent definition can be obtained by defining the time-varying variable $u_t$ that takes the value $1$ if a new update is received at the edge node in time slot $t$, and $0$ otherwise: 
\begin{equation}
    \Delta(t)=\begin{cases}
    \Delta(t-1) +1 & \text{if } u_t=0\\
    1 & \text{if } u_t=1
    \end{cases}
\end{equation}
where $\Delta(0)=0$. This definition of \gls{aoi}, as given in~\cite{kaul2012real}, considers the freshness of information at any given point in time. The long-term expected \gls{aoi} $\Delta_\infty$ is given by:
\begin{equation}
    \Delta_{\infty}=\limsup_{T\rightarrow\infty}\frac{1}{T}\mathbb{E}\left[\sum_{t=1}^{T}\Delta(t)\right].
\end{equation}
This formulation does not include any weighting, assuming that all time steps are equally important. This is reasonable if the monitoring system is either continuous or much faster than the update generating process and communication system, i.e., can be considered as essentially continuous. However, this is only one possibility in real monitoring and control systems: discrete-time systems involve queries in which the monitoring process samples the available information. To capture such applications, we introduce the \gls{qaoi} metric, which samples $\Delta(t)$ according to an arbitrary querying process, thereby considering only the instants at which a query arrives. In this case, we can consider long-term \gls{aoi} as a special case of \gls{qaoi} in which queries arrive at every time instant.

Naturally, in order for an update to be received successfully in slot $t$, we need to transmit it: the \emph{policy} to transmit an update is a function $\pi:\mathcal{S}\rightarrow\{0,1\}$, where $\mathcal{S}$ is a state space and an update is transmitted if the policy outputs 1.

\subsection{The QAoI metric}

If the query arrival process is known in advance, we denote the query arrival times at the edge node by $t_{q,1},t_{q,2},\ldots$. We can then define the \gls{qaoi} for the $i$-th query, denoted as $\tau(i)$ and given by:
\begin{equation}
  \tau(i)=\Delta(t_{q,i}).
\end{equation}
The \gls{eaoi} metric proposed in~\cite{yin2019only} shares many similarities with \gls{qaoi}, and indeed it also represents a pull-based system; however, it deals with concurrent queries and updates differently, allowing the server to wait until the update is over to respond to the query, while our formulation is stricter and enforces an order, with updates always arriving before queries.
We then define the overall objective as minimizing the long-term expected \gls{qaoi}, defined as
\begin{equation}
    \tau_{\infty}=\limsup_{T\to\infty}\frac{1}{T} \mathbb{E}\left[\sum_{i: t_{q,i}\le T} \Delta(t_{q,i})\right].\label{eq:qaoi_delta}
\end{equation}
It is also possible to optimize \gls{qaoi} without full knowledge of future query arrival times, as long as there is some information on the statistics of the process: in our model, the query process is represented by a finite Markov chain with a state space $\mathcal{S}_q$ and a transition matrix $P_q$. The query process is then in a (known) state at any time instant, and queries are generated if the state is in a predetermined subset $\mathcal{Q}\subseteq\mathcal{S}_q$. 

\begin{figure*}[t]
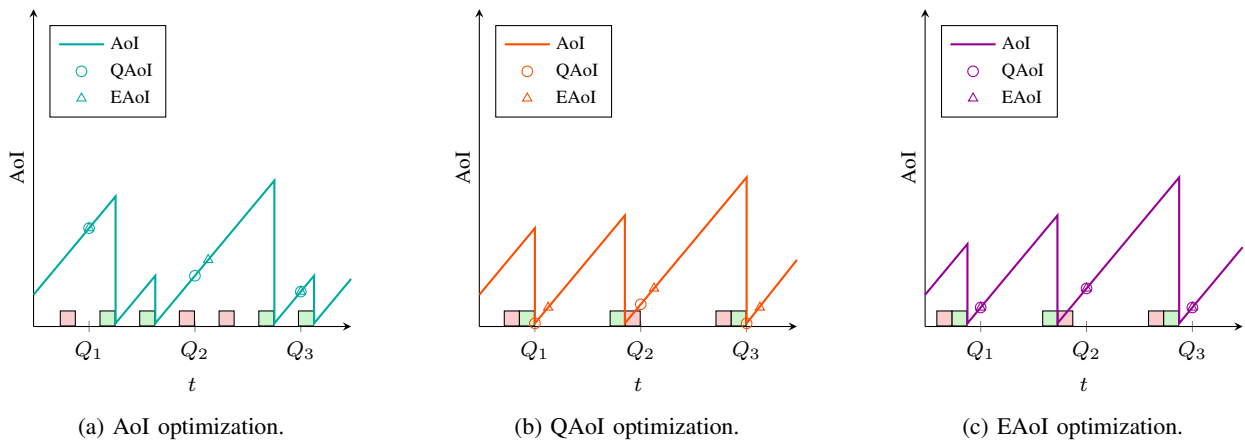

 	\centering
 	\begin{subfigure}[b]{0.32\linewidth}
 		\centering
         \input{JournalFigures/example_qp.tex}
         \caption{AoI optimization.}
         \label{fig:aoi}
     \end{subfigure}	
     \begin{subfigure}[b]{0.32\linewidth}
 		\centering
         \input{JournalFigures/example_qapa.tex}
         \caption{QAoI optimization.}
         \label{fig:qaoi}
     \end{subfigure}     
     \begin{subfigure}[b]{0.32\linewidth}
 		\centering
         \input{JournalFigures/example_eaoi.tex}
         \caption{EAoI optimization.}
         \label{fig:eaoi}
     \end{subfigure}
  \caption{Example of the difference between a system assuming a permanent query and one that is aware of the query arrival process. The same packets are lost in all systems (correctly received packets are depicted in green, lost ones in red), and the markers indicate the age at the query arrival instants.}\vspace{-0.5cm}
  \label{fig:toy_example}
\end{figure*}

Relating this to the use case example, the Markov chain represents the monitoring application: in the simplest case, it requests the sensor reading to the ground station periodically, but in general queries can have complex periodicities that can be modeled by a Markovian process. In most of the simple cases, we have $|\mathcal{Q}|=1$, and the interval between two consecutive queries is \gls{iid}. We can then rewrite the long-term \gls{qaoi} in the more general case with stochastic queries as the following:
\begin{equation}
    %\tau_{\infty}(s)=\limsup_{t\to\infty}\frac{1}{T} \mathbb{E}\left[\sum_{t=1}^T \sum_{s'\in\mathcal{Q}} P_q(s,s') \Delta(t)\right],
    \tau_{\infty}(s_q)=\limsup_{T\to\infty} \mathbb{E}\left[\sum_{t=0}^{T-1}\frac{\Delta(t+1)}{T} \sum_{\mathclap{s_q'(t+1)\in\mathcal{Q}}} P_q(s_q,s_q';t+1) \right],
\end{equation}
where $P_q(s_q,s_q';t+1)$ is the probability of transitioning from state $s_q$ to state $s_q'$ after $t+1$ steps, specified by the $(s_q,s_q')$-th entry in $(P_q)^{t+1}$. In this way, the \gls{qaoi} is considered only in the instants in which a query is happening, i.e., when the Markov chain representing the query process is in a state in which a query arrives. It is also possible to consider normalizing the \gls{qaoi} over the number of queries instead of the number of steps, but this simply leads to a constant multiplying factor as $T$ tends to infinity.

If the query arrival process is memoryless, e.g., queries follow a Bernoulli process, the strategies to minimize \gls{aoi} and \gls{qaoi} are the same, as well as their distributions, as the query process transition probabilities are independent of the current state. The same is true when the query arrival process is much faster than the sensor, i.e., when there is a query in each time slot. The opposite extreme is the case with deterministic query arrivals, in which the transition matrix is deterministic: the query arrival instants are then known \emph{a priori}, and the sensor can optimize its transmissions to minimize \gls{qaoi} directly. The most obvious example of this is given by periodic queries, but the same holds for any deterministic process that is known to the sensor.

In general, the \gls{qaoi} can be very different from the \gls{aoi} in a given system, as well as the strategies for optimizing it. In this work, we present such a class of systems, arguing that an approach that takes the nature and possible periodicity of the monitoring process into account can fit more situations and result in better performance than standard \gls{aoi} minimization. This difference is shown in a simple example in Fig.~\ref{fig:toy_example}, in which the three metrics lead to different solutions: while minimizing \gls{aoi} leads to periodic transmissions, \gls{qaoi} and \gls{eaoi} lead the sensor to transmit just before queries. However, \gls{eaoi}-oriented systems would transmit one slot earlier, so as to avoid delaying the response to the query.

\subsection{Communication system model}\label{sec:comms}

We now define the model of our pull-based communication scenario, in which a sensor needs to schedule transmissions over a link with limited availability. Independently, queries about the state of the sensor are generated, e.g., as part of a monitoring or control process. The objective of this work is to maximize the freshness of the information \emph{at the query time}, while considering that the sensor is energy-constrained and needs to limit the number of transmissions to prolong its lifetime. 

We assume that each update has a fixed size and is transmitted over a \gls{pec} with slotted time. The sensor and receiver are assumed to be synchronized, i.e., the sensor is informed when the last query arrived by the time of the next transmission window.
In order to be as general as possible, we model the channel as a Markov chain with state space $\mathcal{S}_e$ and transition matrix $P_e$, as we did for the query arrival process. An error probability $\epsilon(s)$ is associated to each state $s\in\mathcal{S}_e$, and packets are instantaneously acknowledged by the receiver, so the sensor knows if the last transmitted packet was erased or correctly received. The Markovian model can fit several practical scenarios, including periodic communications or a channel with random outages, and is often used in wireless communications. We also make the simplifying assumption that the sensor can know the state of the channel through beacon messages. This is a simplifying hypothesis, and is not always true in sensor networks. However, having to rely on older estimates of the state of the channel would not change the fundamental nature of the model, and would only increase the uncertainty of the transmission.

The simplest case we can examine is the constant and always available channel, in which $\epsilon(t)=\epsilon$. A slightly more complex example is a deterministic and periodic error process. This models links that are available in a cyclical manner with period $T_e$, like the orbital passes of \gls{leo} communication satellites. In this case, the error probability is $1$ when no satellite is visible and constant during a pass. In our simulations, we limit ourselves to deterministic and periodic error probability processes, whose value is known by the transmitter, but the formulation and solution are general. In our simulations, we consider the four use cases presented in the introduction, which can be simply mapped to different Markov chains for the query and channel error processes: deterministic queries follow a Markov chain with $T_q$ states and forced transmissions, as do predictable channels (with the constant channel as a special case with $T_e=1$). Stochastic query processes and channels follow general finite Markov chains, and in this case, any foresighted action relies on knowledge of the transition probabilities of the Markov model.

\section{Analytical example}\label{ssec:threshold}

\begin{figure*}
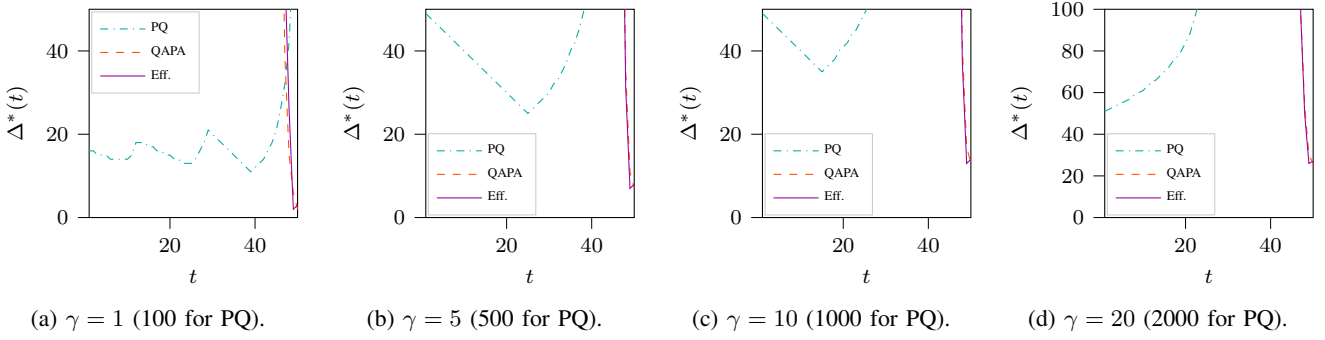

    \centering
    \begin{subfigure}[b]{.24\linewidth}
	    \centering
        \input{RevisionFigures/thr_1.tex}
        \caption{$\gamma=1$ (100 for PQ).}
        \label{fig:th1pq}
    \end{subfigure}	
	\centering
	\begin{subfigure}[b]{.24\linewidth}
	    \centering
        \input{RevisionFigures/thr_5.tex}
        \caption{$\gamma=5$ (500 for PQ).}
        \label{fig:th2pq}
    \end{subfigure}	
    	\begin{subfigure}[b]{.24\linewidth}
	    \centering
        \input{RevisionFigures/thr_10.tex}
        \caption{$\gamma=10$ (1000 for PQ).}
        \label{fig:th3pq}
    \end{subfigure}	
    	\begin{subfigure}[b]{.24\linewidth}
	    \centering
        \input{RevisionFigures/thr_20.tex}
        \caption{$\gamma=20$ (2000 for PQ).}
        \label{fig:th4pq}
    \end{subfigure}	
 \caption{Thresholds $\Delta^*$ for $T=50$, $\varepsilon=0.2$.}\vspace{-0.5cm}
 \label{fig:thr_strategies}
\end{figure*}

We can now consider a simple example in which the difference between \gls{aoi} and \gls{qaoi} is clear, and in which we can derive the optimal strategies analytically. We assume that transmitting has a constant cost $\gamma$, that the channel has a constant error probability $\varepsilon$, and that the optimization is over a finite horizon $T$. The objective of the \gls{pq} system is then to minimize the average \gls{aoi} during an episode, while the \gls{qapa} system will try to minimize the \gls{aoi} at the query instant, i.e., in slot $T$. We also consider the \gls{eaoi} metric from~\cite{yin2019only}, which results in a system similar to \gls{qapa}.

We then define a cost function for the two systems, which we call $c$:
\begin{align}
  c_{\text{PQ}}(t)&=\gamma a_t+\Delta(t);\\
  c_{\text{QAPA}}(t)&=\gamma a_t+\delta(t-T)\Delta(T),\label{eq:cost_qapa}
\end{align}
where $\delta(x)$ is the Dirac delta function. We can also consider a \gls{qapa} system which tries to minimize \gls{eaoi} instead of \gls{qaoi}, to show the difference between the two metrics:
\begin{equation}
  c_{\text{EAoI}}(t)=\gamma a_t+\delta(t-T)\left[\delta(a_t)\Delta(T-1)+(1-\delta(a_t)\Delta(T)\right].
\end{equation}
We then define the long-term cost $C(\tau)=\sum_{t=\tau}^Tc(t),\tau\leq T$. The optimal policy $\pi_{\text{PQ}}^*$ is then the one that minimizes $C_{\text{PQ}}(\tau)$, and the same is true for the \gls{qapa} system. In the last step, with $t=T$, $c(T)=C(T)$, and both systems have the same expected cost if they start from the same conditions and take the same action:
\begin{equation}
  \mathbb{E}[c(T)|a_t=a,\Delta(t-1)=d]=\begin{cases}
                               \gamma+\varepsilon d+1,& \text{if }a=1;\\
                               d+1,& \text{if }a=0.
                             \end{cases}\label{eq:lastreward}
\end{equation}
It is easy to see that transmitting in the last slot, i.e., setting $a_T=1$, reduces the long-term cost if:
\begin{equation}
  \Delta(T-1)>\frac{\gamma}{1-\varepsilon}.\label{eq:lastslot}
\end{equation}
We can then derive the optimal policies. In this case, any policy $\pi(d,t)$ depends only on the current age $\Delta(\tau-1)=d$ and the time slot index $\tau=t$. We only consider cases in which $\epsilon>0$, as the $\epsilon=0$ case is trivially optimized by transmitting only in the last slot before a query. 
\begin{theorem}\label{th:threshold}
  The optimal policies $\pi^*_{\text{PQ}}$ and $\pi^*_{\text{QAPA}}$ are threshold policies, i.e., $\pi_{\text{PQ}}^*(d,t)=1\Rightarrow\pi_{\text{PQ}}^*(d+1,t)=1$ and $\pi_{\text{QAPA}}^*(d,t)=1\Rightarrow\pi_{\text{QAPA}}^*(d+1,t)=1$.
\end{theorem}
The proof of the theorem is given in the Appendix.
The \gls{pq} policy is then defined by threshold values $\Delta_{\text{PQ}}^*(t)$, defined as:
\begin{equation}
\Delta_{\text{PQ}}^*(t)=\text{min}\{d\in\mathbb{N}:\pi_{\text{PQ}}^*(d,t)=1\}.
\end{equation}
The same holds for the \gls{qapa} policy.
 The threshold for transmission can be computed recursively: if we know $\pi_{\text{PQ}}^*$ from time $\tau+1$ to the end of the episode, we can compute $\Delta_{\text{PQ}}^*(\tau)$ using the following formula:
\begin{equation}
\begin{aligned}
  \Delta_{\text{PQ}}^*(\tau)=\text{inf}\Big\{&d\in\mathbb{N}:d+\mathbb{E}[C(\tau+1)|\pi^*,\Delta(\tau)=d]\\
  &>\frac{\gamma}{1-\epsilon}+\mathbb{E}[C(\tau+1)|\pi^*,\Delta(\tau)=0]\!\Big\},
\end{aligned}
\end{equation}
where $\mathbb{E}[C(\tau+1)|\pi^*,\Delta(\tau)=d]$ is the expected long-term cost while following the optimal policy. On the other hand, $\Delta_{\text{QAPA}}^*(\tau)$ is given by:
\begin{equation}
\begin{aligned}
  \Delta_{\text{QAPA}}^*(\tau)=\text{inf}\Big\{&d\in\mathbb{N}:\mathbb{E}[C(\tau+1)|\pi^*,\Delta(\tau)=d]\\
  &>\frac{\gamma}{1-\epsilon}+\mathbb{E}[C(\tau+1)|\pi^*,\Delta(\tau)=0]\Big\}.
\end{aligned}
\end{equation}
Naturally, we have $\Delta_{\text{PQ}}^*(T)=\Delta_{\text{QAPA}}^*(T)=\frac{\gamma}{1-\epsilon}$, as given by~\eqref{eq:lastslot}.

Fig.~\ref{fig:thr_strategies} shows how the threshold strategies work in a system with $\varepsilon=0.2$: while the \gls{pq} system exhibits some periodic behavior, mostly due to the effect of the finite horizon, it tends to transmit less if the episode is close to the end, as the future reduction of the \gls{aoi} is limited in time. On the other hand, the \gls{qapa} sensor transmits only in the last steps before the query, but increasing the value of $\gamma$ makes it less convenient to transmit, and the sensor will only do so for a progressively higher expected \gls{qaoi}, and only closer to the actual query instant. If we use \gls{eaoi} as a metric for the \gls{qapa} system, the threshold is not monotonic any more: the best moment to transmit is actually one step before the query, as transmitting in the same slot increases the delay in the query response. The transmission cost $\gamma$ for \gls{pq} is higher than for both \gls{qapa} settings, because the overall cost is much higher, taking into account the \gls{aoi} at every step and not just at the last step. The monotonicity of the strategy for \gls{qaoi}, which holds over all our simulations in more complex systems, can make optimization easier, and a formulation without delayed responses can be simpler to implement for \gls{iot} gateways.

\begin{figure}
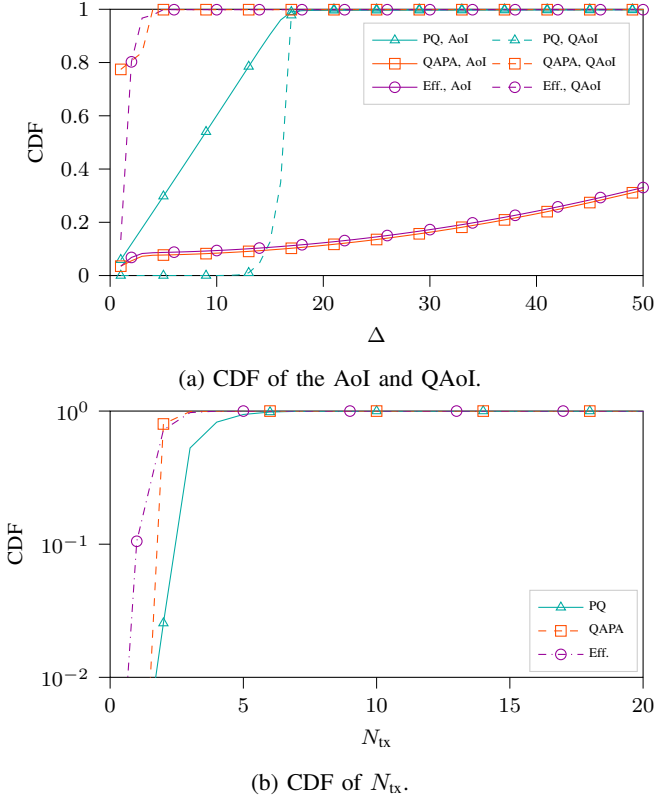

    \centering
    	\begin{subfigure}[b]{\columnwidth}
	    \flushright
    \input{RevisionFigures/thr_cdf1}
        \caption{\gls{cdf} of the \gls{aoi} and \gls{qaoi}.}
        \label{fig:thr_CDF}
    \end{subfigure}	
    	\begin{subfigure}[b]{\columnwidth}
	    	    \flushright
        \input{RevisionFigures/tx_cdf1}
        \caption{\gls{cdf} of $N_{\text{tx}}$.}
        \label{fig:thr_saw}
    \end{subfigure}	
    \caption{Behavior of the \gls{pq}, \gls{qapa}, and \gls{eaoi} systems with $T=50$ and $\varepsilon=0.2$, starting from a random age $\Delta_0$ between 1 and 100.}\vspace{-0.5cm}
    \label{fig:threshold_mc}
\end{figure}

The performance of the systems is shown in Fig.~\ref{fig:threshold_mc}, which shows the \gls{cdf} of the \gls{aoi} and \gls{qaoi}, along with the number of transmissions in each episode $N_{\text{tx}}$: as we can see, the \gls{qapa} strategy has a much better \gls{qaoi}, even though it transmits slightly less. The \gls{eaoi}-based system has a slightly higher \gls{qaoi} than the \gls{qaoi}-oriented system, as it tends to transmit slightly earlier: however, it also has slightly fewer transmission attempts on average. For both systems, this comes at the cost of the \gls{aoi}, which is significantly higher than for the \gls{pq} system. As we will see in the next sections, this basic pattern holds even for much more complex systems.

\section{MDP formulation and problem solution}\label{sec:mdp}
In this section, we will consider the full problem, with a query process and energy dynamics. To understand the impact of the query process in the performance of a communication system, we will model the \gls{pq} and \gls{qapa} systems
as \glspl{mdp}, which we will then proceed to solve. 
A \gls{mdp} is defined by a state space 
$\mathcal{S}$, an action space $\mathcal{A}$, a set of transition probabilities $p_a(s,s')=P(s_{t+1}=s'|a_t=a,s_t=s)$,  and an instantaneous reward function $r(s,a,s')$, which represents the immediate reward when taking action $a$ and transitioning from state $s$ to state $s'$. 
The two systems, \gls{pq} and \gls{qapa}, can use the same state and action spaces, and only differ in the reward function that they use. This problem formulation is significantly more complex than the example from Sec.~\ref{ssec:threshold}, making the direct derivation of strategies complex, but it can represent a more general class of communication systems.

To model the energy-constrained nature of the node, we use a \emph{leaky bucket} model, as commonly done in the literature~\cite{raghunathan2004energy}: we consider a bucket of tokens, which is replenished by a process which can generate tokens independently at each step with probability $\mu_b$. The node can only transmit a packet if there are tokens in the bucket, and each transmission consumes one token. This model can fit a general power consumption constraint on a battery-powered node, which should limit its number of transmissions in order to prolong its lifetime. Furthermore, it allows us to easily include the constraint in the \gls{mdp} formulation as elaborated further in the next section.

Decisions are made by the sensor at every slot, as it can either keep silent or send a packet. Consequently, the action space is $\mathcal{A}=\{0,1\}$. Both the \gls{pq} and the \gls{qapa} agents (i.e., sensors with two different objectives) need to know the current age $\Delta(t)$, as well as the state $s_e(t)\in\mathcal{S}_e$ of the error probability Markov process. Additionally, the agent should know the number of available tokens, $b(t)$, as it will influence its decision whether to transmit. If the number of tokens is 0, the sensor is blocked from transmitting until a token is generated. The tuple $(\Delta(t), s_e(t), b(t))$ is sufficient to represent the state in the \gls{pq} system, which does not require any knowledge of the query arrival process, while the \gls{qapa} system adds a fourth element to the state, i.e., the state of the query arrival process $s_q(t)\in\mathcal{S}_q$. As the \gls{pq} system can be studied as a special case of the \gls{qapa} system (with a single-state query arrival process), we adopt the wider definition for both systems to simplify the notation. We then define the state space as $\mathcal{S}=\mathbb{N}^2\times\mathcal{S}_e\times\mathcal{S}_q$
and assume that the query arrival, token generation, and error probability processes are independent, examining each element of the state separately.
The \gls{aoi} increases by one between each slot unless the node decides to transmit and the packet is successfully received, with probability $p_s(t)=1-\epsilon(s_e(t))$, in which case the \gls{aoi} is reduced to one in the subsequent slot. The transition probabilities are thus described by
\begin{align}
    &P(\Delta(t+1)=d|s_t,a_t) = \begin{cases} a_tp_s(t) &d=1;\\
    1-a_tp_s(t) &d=\Delta(t)+1;
    \\
    0 &\text{otherwise,}
    \end{cases}
\end{align}
where $a_t$ is the action at time $t$, which equals zero if the sensor is silent and one if it transmits.
Secondly, the number of tokens in the next slot depends on whether a new token is generated and whether the sensor transmits, in which case it uses one token. The transition probability from $b(t)$ to $b(t+1)$ is:
\begin{equation}
P(b(t+1)=b+i|b(t),a_t)=\begin{cases}
 \mu_b&\text{if }i=1-a_t;\\
 1-\mu_b&\text{if }i=-a_t;\\
 0 &\text{otherwise.}
\end{cases}
\end{equation}
The transition probabilities for the error probability and query arrival processes are defined by the matrices $P_e$ and $P_q$, respectively. We assume that the query process is error-free, i.e., that the link between the ground station and the monitor is error-free.

We define two cost functions; one for the \gls{pq} system, which does not depend on the query instant and will be used as baseline, and one for the \gls{qapa} system, in which the cost is only considered when a query arrives. In the baseline \gls{pq} model, the cost is given by the \gls{aoi} in any slot:
\begin{equation}\label{eq:cost_AoI}
    c_{\text{PQ}}(s_t,a_t,s_{t+1})=\Delta(t+1).
\end{equation}
However, in the \gls{qapa} system, the cost is the \gls{aoi} when a query arrives:
\begin{equation}\label{eq:cost_QAoI}
    c_{\text{QAPA}}(s_t,a_t,s_{t+1})=\begin{cases}
    \Delta(t+1) &\text{if }s_q(t+1)\in\mathcal{Q};\\
    0 &\text{otherwise.}
    \end{cases}
\end{equation}
In both cases, the objective is to find a policy $\pi^*$ that minimizes the long-term cost. In this initial work, we limit ourselves to consider the discounted case, which benefits from strong convergence guarantees, and defer the case with undiscounted costs to future work. In this case, at least one optimal policy is guaranteed to exist as a stationary deterministic decision rule~\cite{bellman1957markovian}, i.e. $\pi^*: \mathcal{S}\to\mathcal{A}$. Specifically, we solve
\begin{equation}
\pi^*=\arg\min_{\pi}\left[\sum_{t=0}^\infty \lambda^t\sum_{s_t\in\mathcal{S}} P(s_t|s_0,\pi) c(s_t,\pi(s_t))\right],
\end{equation}
where $\lambda< 1$ is the discount factor, and $c(s_t,a_t)=\mathbb{E}_{s_{t+1}}[c(s_t,a_t,s_{t+1})|s_t,a_t]$ is the expected cost of taking action $a_t$ in state $s_t$ under either the PQ or QAPA model. Naturally, the energy constraint has a major impact on the cost, but it is implicit: as the sensor cannot transmit an update if it has no energy tokens, its age will increase, consequently increasing the long-term cost. In fact, we explicitly excluded a cost of transmission from the model, as the policy can already account for energy limitations by tuning its behavior and avoiding short-sighted choices that maximize the short-term reward while leading to long-term harm. The factor $\lambda$ is crucial in this, as a higher discount value leads to a more foresighted policy.

\subsection{Problem solution}

We can now proceed to solve the \gls{mdp} for the two systems we have defined using policy iteration, as described in \cite[Ch. 4]{sutton2018reinforcement}. In order to apply the algorithm, we need to truncate the problem to a finite \gls{mdp}. We do so by defining a maximum age $\Delta_{\max}$, a maximum query interval $T_{q,\max}$, and a token bucket size $B$: once the age, the query interval, or the number of tokens in the bucket reach the maximum, they cannot increase further. As long as the maximum values are sufficiently large, they are not reached during normal operation and this simplification does not affect the optimal policies or their performance.

The policy iteration algorithm has two steps, policy evaluation and policy improvement, which are repeated until convergence. The algorithm is initialized with a policy function $\pi^0$ and a value function $v_\pi^0$, which are both set to all zeros. At each step $n$, we can use the current estimate of the value of a state to update the next estimate, getting the value $u^n(s,a,s')$, defined as:
\begin{equation}
 u^n(s,a,s')=c(s,\pi^n(s),s')+\lambda v^n_\pi(s').
\end{equation}

The iterative steps are then:
\begin{enumerate}
    \item The policy is evaluated using \begin{equation}
        v_\pi^{n+1}(s)=\sum_{s'\in\mathcal{S}}p(s'\vert s,\pi^n(s))u^n(s,a,s'),
    \end{equation}
     for all $s$, where $s$ is the current state, $s'$ is the new state, $a$ is the action, and $c$ is the cost from either \eqref{eq:cost_AoI} or \eqref{eq:cost_QAoI}. The value function is an estimate of the long-term value that can be achieved in a given state using the policy.
    \item The policy is improved by choosing the action that maximizes the long-term value, i.e., minimizes the long-term cost:
    \begin{equation}
        \pi^{n+1}(s)=\argmin_{a\in\mathcal{A}} \sum_{s'\in\mathcal{S}}p(s'\vert s,a)u^{n+1}(s,a,s').
    \end{equation}
\end{enumerate}
Policy iteration is guaranteed to converge to the optimal policy~\cite{howard1960dynamic} in finite-state \glspl{mdp} with finite reward. The complexity of policy iteration in general is exponential in the number of states, making it particularly impractical for realistic problems. However, if the correct pivoting rule is adopted and the discount factor is constant in time, policy iteration was shown to be strongly polynomial in~\cite{ye2011simplex}:
\begin{equation}
  N_{\text{it}}\leq\frac{|\mathcal{S}|^2(|\mathcal{A}|-1)}{1-\lambda}\log\left(\frac{|\mathcal{S}|^2}{1-\lambda}\right).\label{eq:comput}
\end{equation}
Each iteration uses at most $O(|\mathcal{A}||\mathcal{S}|^2)$ operations, making the resulting bound polynomial in the size of the \gls{mdp}.
As mentioned above, we truncated the age and token bucket size to make the \gls{mdp} finite, so the conditions to use the algorithm apply. The notation in the past two sections is summarized in Table~\ref{tab:notation}. In our case, in which there are only two actions, policy iteration then converges in $O\left(\frac{|\mathcal{S}|^2}{1-\lambda}\log\left(\frac{|\mathcal{S}|^2}{1-\lambda}\right)\right))$ iterations, which correspond to $O\left(\frac{2|\mathcal{S}|^4}{1-\lambda}\log\left(\frac{|\mathcal{S}|^2}{1-\lambda}\right)\right)$ steps.

Naturally, policy iteration can suffer from the curse of dimensionality in more complex scenarios: as the number of states increases exponentially with the number of considered dimensions of the state, and policy iteration has itself an exponentially increasing complexity as the number of states increases, this solution is only practical for relatively small problems. In larger problems, learning-based techniques such as reinforcement learning can provide a far faster convergence, particularly when using deep reinforcement methods that can generalize experience~\cite{mnih2015human}. These solutions have been successfully applied in the \gls{aoi} optimization literature~\cite{ceran2019reinforcement}, and they can be applied directly to our \gls{mdp} formulation, but we consider them as part of future work on this subject, solving some simpler examples with policy iteration.

\section{Simulation settings and results}\label{sec:results}

\begin{figure*}
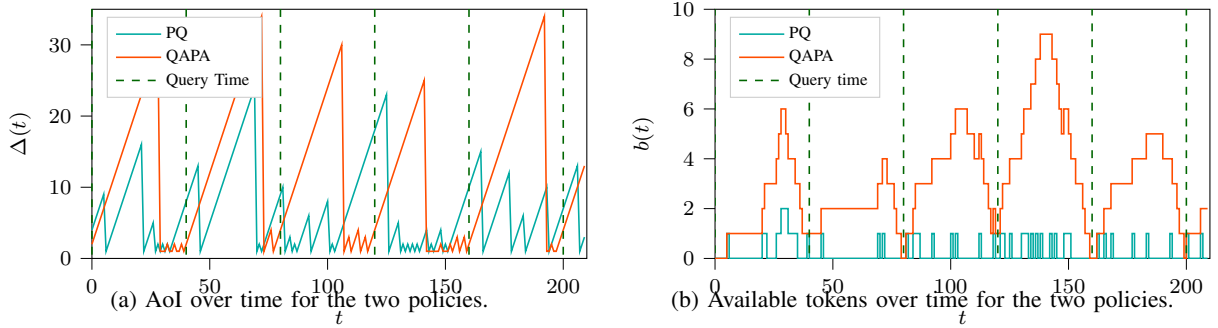

    \centering
    	\begin{subfigure}[b]{.45\linewidth}
	    \centering
    \input{JournalFigures/sawtooth}\vspace{-0.8cm}
        \caption{\gls{aoi} over time for the two policies.}
        \label{fig:sawtooth}
    \end{subfigure}	
    	\begin{subfigure}[b]{.45\linewidth}
	    \centering
        \input{JournalFigures/sawtooth_tokens}\vspace{-0.8cm}
        \caption{Available tokens over time for the two policies.}
        \label{fig:tokens}
    \end{subfigure}	
    \caption{\Gls{aoi} dynamics of the \gls{pq} and \gls{qapa} policies for $T_q=40,\mu_b=0.2,\epsilon=0.2$. The \Gls{pq} policy generally has a lower \gls{aoi}, but the \gls{qapa} policy minimizes the \gls{aoi} at the query instants.}\vspace{-0.5cm}
    \label{fig:Sawtooth1}
\end{figure*}

This section presents Monte Carlo evaluations of the policies obtained using the \gls{mdp} described in Section~\ref{sec:mdp}. Although, the methods in Section~\ref{sec:mdp} can be applied to any query process, throughout the evaluation we will consider queries that occur periodically, at a fixed time interval $T_q$. Furthermore, we truncate the \gls{mdp} at a maximum age of $\Delta_{\text{max}}=100\times T_q$ and a maximum token bucket size of $B=10$, and we use a discount factor $\lambda=0.95$. We use the term \gls{aoi} to refer to the age at any time and \gls{qaoi} for the age sampled at the query instants.

\subsection{Periodic queries with constant error probability}

We first consider the simplest scenario, in which the error probability is constant and the query arrival process is deterministic with period $T_q$. In this scenario, the error probability process only has one state, i.e., $|\mathcal{S}_e|=1$, and the error probability is a constant value $\epsilon$. The query arrival process is a deterministic Markov chain with $T_q$ states, with $\mathcal{S}_q=\{1,\ldots,T_q\}$. The transition probabilities are given by:
\begin{equation}
P_q(s_q,s_q')=\begin{cases}
               1, & s_q<T_q\wedge s_q'=s_q+1;\\
               1, & s_q=T_q\wedge s_q'=1;\\
               0, &\text{otherwise.}
              \end{cases}\label{eq:det_queries}
\end{equation}
The subset of query states is given by $\mathcal{Q}={T_q}$, i.e., the server sends a query only when the state reaches $T_q$, after which the state is reset to 1 and the counter starts increasing again. This Markov chain is equivalent to a periodic deterministic query process.

\begin{figure*}
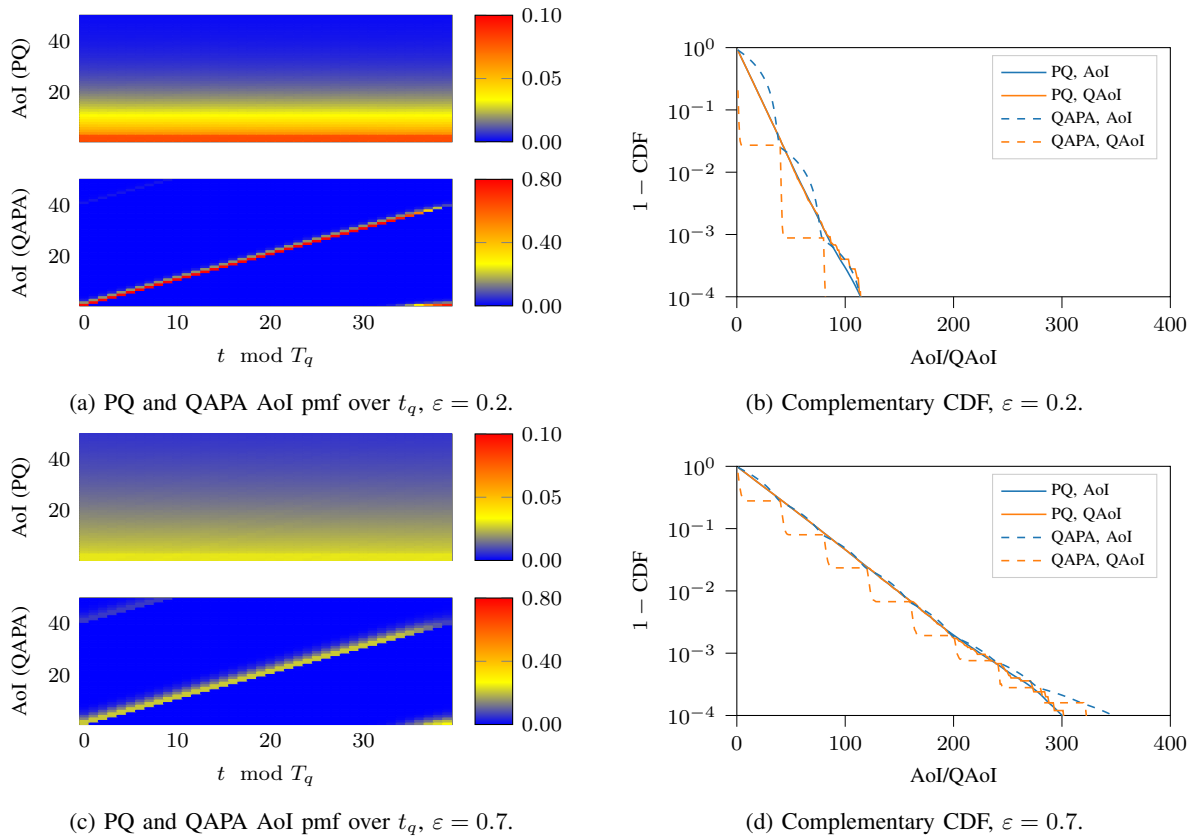

    \centering
	\begin{subfigure}[b]{.45\linewidth}
	    \centering
	    %\tikzexternalenable
        \input{RevisionFigures/matrix_02_case1}
        %\tikzexternaldisable
        \caption{\gls{pq} and \gls{qapa} AoI pmf over $t_q$, $\epsilon=0.2$.}
        \label{fig:matrix_02}
    \end{subfigure}	
	\centering
	\begin{subfigure}[b]{.45\linewidth}
	    \centering
        \input{RevisionFigures/CDF,error=0.2,Tq=40,Pb=0.1,X=10,case1}
        \caption{Complementary CDF, $\epsilon=0.2$.}
        \label{fig:CDF_0.2_40_01}
    \end{subfigure}	
	\begin{subfigure}[b]{.45\linewidth}
	    \centering
	    %\tikzexternalenable
        \input{RevisionFigures/matrix_07_case1}
        %\tikzexternaldisable
        \caption{\gls{pq} and \gls{qapa} AoI pmf over $t_q$, $\epsilon=0.7$.}
        \label{fig:matrix_07}
    \end{subfigure}	
    \begin{subfigure}[b]{.45\linewidth}
	    \centering
        \input{RevisionFigures/CDF,error=0.7,Tq=40,Pb=0.1,X=10,case1}     
        \caption{Complementary CDF, $\epsilon=0.7$.}
        \label{fig:CDF_0.7_40_01}
    \end{subfigure}
     \caption{AoI distributions and CCDFs for \gls{pq} and \gls{qapa} for $T_q=40,\mu_b=0.1$, and $\epsilon=\{0.2,0.7\}$.}\vspace{-0.5cm}
 \label{fig:CDF}
\end{figure*}

We start by exploring the temporal dynamics of the \gls{aoi} process obtained using the \gls{pq} and the \gls{qapa} policies. Recall that \gls{pq} is optimized to achieve a low \gls{aoi} independent of the query process, while \gls{qapa} minimizes the \gls{aoi} at the query times, using cost functions \eqref{eq:cost_AoI} and \eqref{eq:cost_QAoI}, respectively. Fig.~\ref{fig:sawtooth} shows the \gls{aoi} for queries occurring periodically every $T_q=40$ time slots as indicated by the vertical lines, a packet error probability of $\epsilon=0.2$, and a token rate $\mu_b=0.2$. It is seen that the \gls{pq} policy reduces the \gls{aoi} approximately uniformly across time, while the \gls{qapa} policy consistently tries to reduce the \gls{aoi} in the slots immediately prior to a query, so that the \gls{aoi} is minimized when the query arrives. This is reflected in Fig.~\ref{fig:tokens}, which shows that the \gls{qapa} policy accumulates energy when the next query is far in the future, unlike \gls{pq}. A consequence of this is that the \gls{qapa} policy generally has a slightly higher average \gls{aoi} than the \gls{pq} policy, but the \gls{qaoi} of the \gls{qapa} is significantly lower than that of the \gls{pq} policy.

The initial observations from Fig.~\ref{fig:Sawtooth1} can be confirmed by the distribution of the \gls{aoi} as a function of the time since the last query, as illustrated in Fig.~\ref{fig:CDF}. Figs.~\ref{fig:CDF}\subref{fig:matrix_02} and \ref{fig:CDF}\subref{fig:matrix_07} show the \gls{pmf} of the \gls{aoi} conditioned on various time instants $t\mod T_q$, while Figs.~\ref{fig:CDF}\subref{fig:CDF_0.2_40_01} and \ref{fig:CDF}\subref{fig:CDF_0.7_40_01} show the \gls{cdf} of the overall \gls{aoi} and \gls{qaoi}. We can immediately see the difference between the two policies from Fig.~\ref{fig:CDF}\subref{fig:matrix_02}: the horizontal axis represents the time since the last query, while the colors represent the \gls{pmf} of the \gls{aoi}, whose domain is on the vertical axis. The \gls{aoi} for the \gls{pq} policy does not depend on the time since the last query: the distribution is the same for all time instants, as can be seen by the fact that each horizontal line in the plot has exactly the same color. On the other hand, the \gls{qapa} policy shows a very different pattern: the \gls{aoi} increases linearly as time passes, which indicates that the sensor does not send any packets in the first half of the interval, then sharply drops and stays very close to 0 in the final part of the interval. This behavior is consistent with what we would expect from a \gls{qaoi}-oriented system, as sending packets long before the next query is basically a waste of energy, and transmissions are clustered just before the query instant. The resulting \gls{cdf} in Fig.~\ref{fig:CDF}\subref{fig:CDF_0.2_40_01} reveals, as expected, that the \gls{aoi} and the \gls{qaoi} are equivalent for the \gls{pq} policy, as the distribution is the same at any time instant. However, for the \gls{qapa} policy, the \gls{qaoi} is significantly lower than the \gls{aoi}, while the \gls{aoi} is often larger than the \gls{pq} policy's. This is due to the fact that the \gls{qaoi} is only measured at the query instants, at which the age of the \gls{qapa} policy is minimized. Due to the energy constraint, this comes at the cost of a generally higher age, causing a higher \gls{aoi} measured at each time instant. Finally, the staircase appearance in the \gls{cdf} is due to the fact that the queries happen periodically. If the queries were arriving at variable (but known in advance) intervals, then the \gls{qapa} would still have lower \gls{qaoi} than the \gls{pq} query, but its \gls{cdf} would have a different shape.

The same observations apply for the scenario with high error probability, $\epsilon=0.7$, shown in Figs.~\ref{fig:CDF}\subref{fig:matrix_07} and \ref{fig:CDF}\subref{fig:CDF_0.7_40_01}. Although the \gls{aoi} and \gls{qaoi} 
are higher due to the high packet error rate, the applied policies are similar. It is interesting to note that there is a significant probability of having a \gls{qaoi} higher than 40 (which corresponds to $T_q$) in Fig.~\ref{fig:CDF}\subref{fig:matrix_07}, and that transmissions tend to be even more clustered towards the end of the query interval for the \gls{qapa} policy, indicating that the sensor will tend to save energy for future queries when it gets several tokens. Although there is a significant probability that the packet immediately prior to the query is lost, the \gls{aoi} distribution at $t\mod T_q=0$ is still concentrated close to one.

\begin{figure*}[t!]
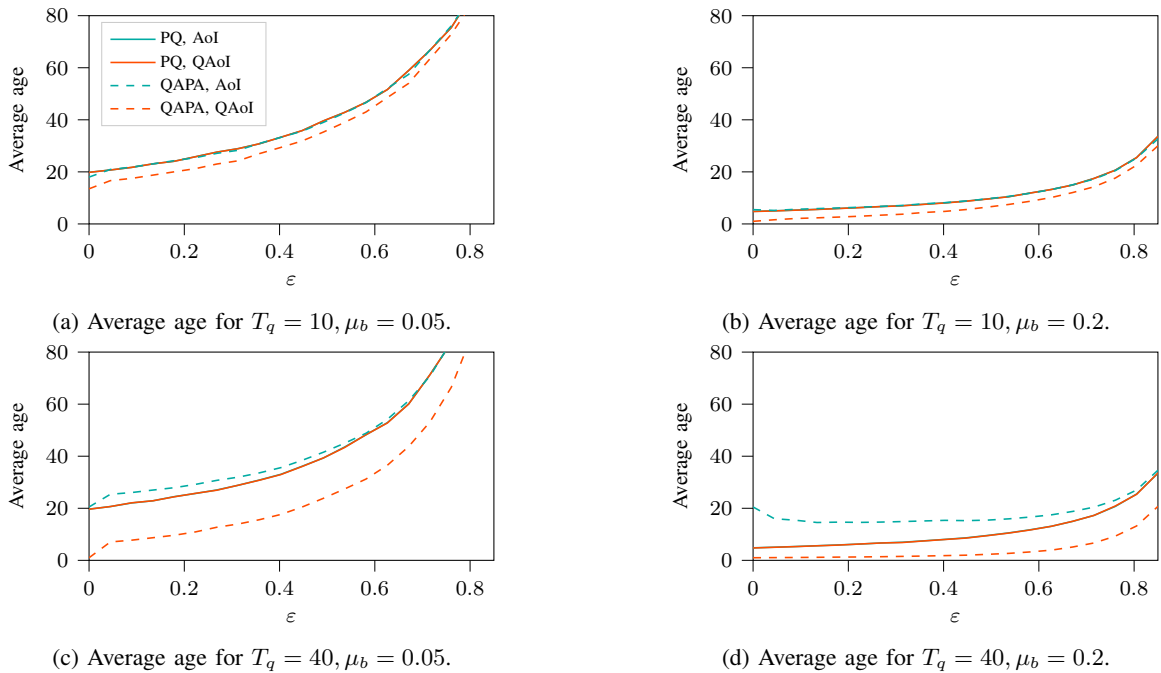

	\centering
	\begin{subfigure}[b]{0.48\linewidth}
	    \centering
        \input{RevisionFigures/AverageAge,Tq=10,Pb=0.05,X=10,case1}
        \caption{Average age for $T_q=10,\mu_b=0.05$.}
        \label{fig:avg_10_005}
    \end{subfigure}		
    \begin{subfigure}[b]{0.48\linewidth}
	    \centering
        \input{RevisionFigures/AverageAge,Tq=10,Pb=0.2,X=10,case1}    
        \caption{Average age for $T_q=10,\mu_b=0.2$.}
        \label{fig:avg_10_02}
    \end{subfigure}	
%     \begin{subfigure}[b]{0.48\linewidth}
% 	    \centering
% %        \input{Figures/AverageAge,Tq=20,Pb=0.05}
%         \input{RevisionFigures/AverageAge,Tq=20,Pb=0.05,X=10,case1}       
%         \caption{Average age for $T_q=20,\mu_b=0.05$.}
%         \label{fig:avg_20_005}
%     \end{subfigure}
%     \begin{subfigure}[b]{0.48\linewidth}
% 	    \centering
% %        \input{Figures/AverageAge,Tq=20,Pb=0.2} 
%         \input{RevisionFigures/AverageAge,Tq=20,Pb=0.2,X=10,case1}       
%         \caption{Average age for $T_q=20,\mu_b=0.2$.}
%         \label{fig:avg_20_02}
%     \end{subfigure}
	\begin{subfigure}[b]{0.48\linewidth}
	    \centering
        \input{RevisionFigures/AverageAge,Tq=40,Pb=0.05,X=10,case1}      
        \caption{Average age for $T_q=40,\mu_b=0.05$.}
        \label{fig:avg_40_005}
	\end{subfigure}
	\begin{subfigure}[b]{0.48\linewidth}
	    \centering
        \input{RevisionFigures/AverageAge,Tq=40,Pb=0.2,X=10,case1}    
        \caption{Average age for $T_q=40,\mu_b=0.2$.}
        \label{fig:avg_40_02}
	\end{subfigure}
 \caption{Average \gls{aoi} and \gls{qaoi} for the two systems for different values of $T_q$ and $\mu_b$.}\vspace{-0.5cm}
 \label{fig:average_age}
\end{figure*}

We now study how the average \gls{aoi} and \gls{qaoi} changes with the packet error probability $\epsilon$ for various choices of the parameters, shown in Fig.~\ref{fig:average_age}. For all cases, the \gls{qapa} policy achieves the lowest \gls{qaoi}, while the \gls{pq} policy achieves the lowest \gls{aoi}. When the query period, $T_q$, is low, the difference between \gls{aoi} and \gls{qaoi} is relatively small, as is the difference between the two policies. Intuitively, this is because the query instants, which are prioritized by the \gls{qapa} policy, are more frequent, making the two problems more similar.

\begin{figure*}
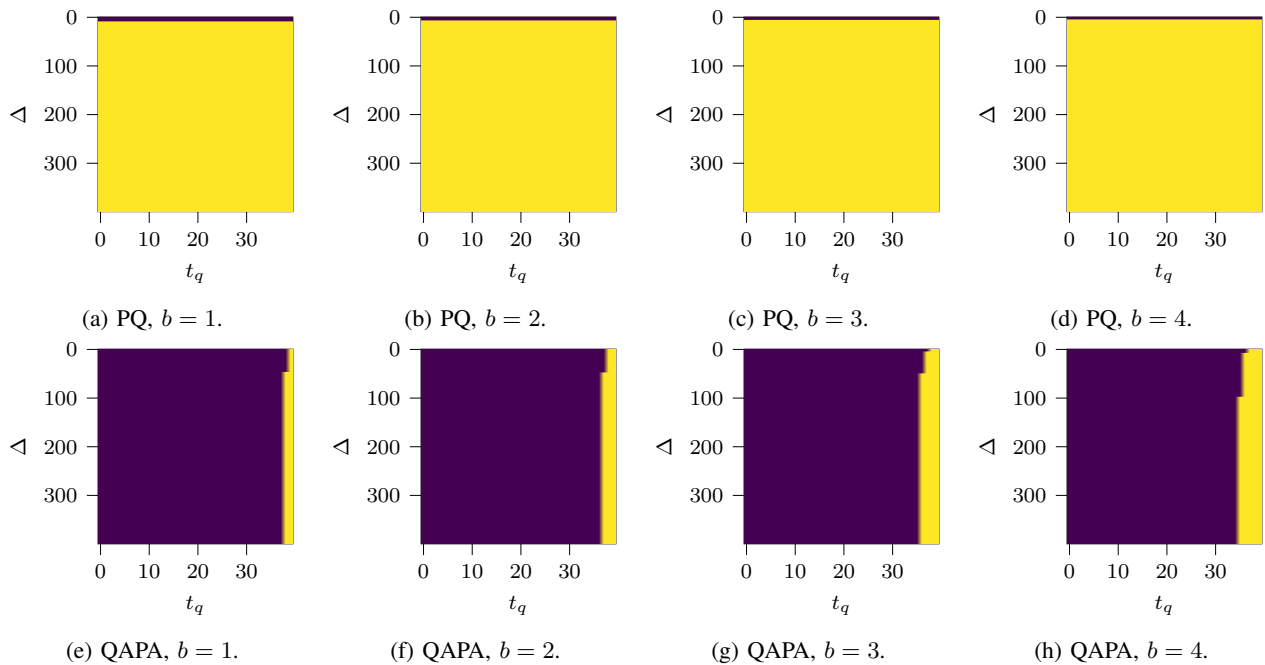

    \centering
    \begin{subfigure}[b]{.23\linewidth}
	    \centering
        \input{RevisionFigures/pq_strategies,b=1.tex}
        \caption{\gls{pq}, $b=1$.}
        \label{fig:pq1}
    \end{subfigure}	
	\centering
	\begin{subfigure}[b]{.23\linewidth}
	    \centering
        \input{RevisionFigures/pq_strategies,b=2.tex}
        \caption{\gls{pq}, $b=2$.}
        \label{fig:pq2}
    \end{subfigure}	
    	\begin{subfigure}[b]{.23\linewidth}
	    \centering
        \input{RevisionFigures/pq_strategies,b=3.tex}
        \caption{\gls{pq}, $b=3$.}
        \label{fig:pq3}
    \end{subfigure}	
    	\begin{subfigure}[b]{.23\linewidth}
	    \centering
        \input{RevisionFigures/pq_strategies,b=4.tex}
        \caption{\gls{pq}, $b=4$.}
        \label{fig:pq4}
    \end{subfigure}	
	\begin{subfigure}[b]{.23\linewidth}
	    \centering
        \input{RevisionFigures/qapa_strategies,b=1.tex}
        \caption{\gls{qapa}, $b=1$.}
        \label{fig:qapa1}
    \end{subfigure}	
	\centering
	\begin{subfigure}[b]{.23\linewidth}
	    \centering
        \input{RevisionFigures/qapa_strategies,b=2.tex}
        \caption{\gls{qapa}, $b=2$.}
        \label{fig:qapa2}
    \end{subfigure}	
    	\begin{subfigure}[b]{.23\linewidth}
	    \centering
        \input{RevisionFigures/qapa_strategies,b=3.tex}
        \caption{\gls{qapa}, $b=3$.}
        \label{fig:qapa3}
    \end{subfigure}	
    	\begin{subfigure}[b]{.23\linewidth}
	    \centering
        \input{RevisionFigures/qapa_strategies,b=4.tex}
        \caption{\gls{qapa}, $b=4$.}
        \label{fig:qapa4}
    \end{subfigure}	
 \caption{Strategies for the two systems with $\varepsilon=0.2$. Yellow indicates a transmission, dark blue indicates no transmission.}\vspace{-0.5cm}
 \label{fig:strategies}\end{figure*}

As a result, awareness of the query arrival process becomes more important when queries are rare, i.e., when $T_q$ is large: this is clear from the large gap between the average \gls{qaoi} achieved by \gls{qapa} and by \gls{pq} in Fig.~\ref{fig:avg_40_005} and Fig.~\ref{fig:avg_40_02}.
The plots on the left show the results for $\mu_b=0.05$, i.e., when a new token is generated on average every 20 time slots. When $T_q=10$ (see Fig.~\ref{fig:average_age}\subref{fig:avg_10_005}), the token period becomes a limiting factor, and both the \gls{aoi} and \gls{qaoi} are relatively high even for low values of $\epsilon$. 
In particular, in the error-free case when $\epsilon=0$, the average \gls{qaoi} cannot be lower than $(1+11)/2=6$, which is achieved by transmitting an update prior to every second query. Interestingly, the impact of the energy limit becomes less significant for the \gls{qapa} policy as the time between queries increases: by saving up tokens until right before the query, this policy can significantly reduce the \gls{qaoi}, at the cost of a higher \gls{aoi}. On the other hand, the \gls{pq} policy does not benefit from this increase, as it is oblivious of the query arrival frequency. When tokens are generated faster, at rate $\mu_b=0.2$, as shown in the plots on the right side of the figure, the \gls{aoi} and the \gls{qaoi} are generally lower, since more frequent transmissions are allowed.

Finally, it is interesting to look at the strategies for the two systems, which are plotted in Fig.~\ref{fig:strategies}. As expected, the optimal strategy for the \gls{pq} system does not depend on the queries, only on the current \gls{aoi}. If the \gls{aoi} is very low, the \gls{pq} system waits for a while to transmit, but this is limited to a few slots (8 if $b=1$, and just 4 if $b=4$). On the other hand, while the \gls{pq} strategy plots are horizontal, the \gls{qapa} plots are almost vertical, i.e., the importance of the \gls{aoi} is very low, and the system decides almost only based on the time until the next query. Naturally, if $b$ is low, it only tries to transmit in the very last possible slot, to reduce the \gls{qaoi}, while more tokens allow it to start earlier and protect itself from errors. If the \gls{aoi} is high enough, it might be worth it to start transmitting a little earlier and hope to get a new token in the next slot, as the small increase in the \gls{qaoi} if only the packet transmission in the earlier slot is successful is offset by the large decrease if the new token arrives and only that transmission is successful. If we compare the \gls{qapa} strategies with the threshold-based ones derived for the simplified system in Fig.~\ref{fig:thr_strategies}, we see a significant similarity: the basic structure is still the same, although the precise values of the thresholds change. This result is encouraging for future attempts at deriving analytical threshold-based strategies for the full problem.

\begin{figure*}
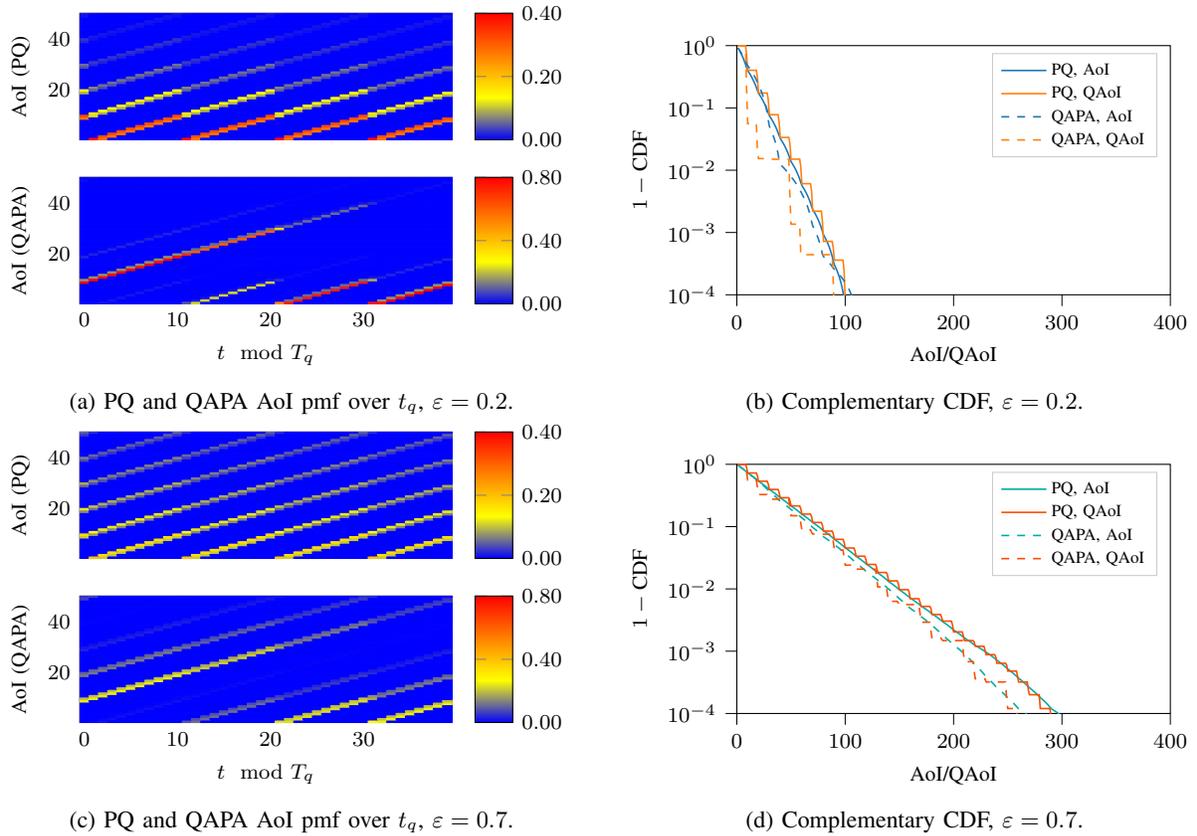

    \centering
	\begin{subfigure}[b]{.45\linewidth}
	    \centering
        \input{RevisionFigures/matrix_02_case2}
        \caption{\gls{pq} and \gls{qapa} AoI \gls{pmf} over $t_q$, $\epsilon=0.2$.}
        \label{fig:matrix_02_te}
    \end{subfigure}
	\centering
	\begin{subfigure}[b]{.45\linewidth}
	    \centering
		\input{RevisionFigures/CDF,error=0.2,Tq=40,Pb=0.1,X=10,case2}     
        \caption{Complementary CDF, $\epsilon=0.2$.}
        \label{fig:CDF_0.2_40_01_te10}
    \end{subfigure}	
	\begin{subfigure}[b]{.45\linewidth}
	    \centering
        \input{RevisionFigures/matrix_07_case2}
        \caption{\gls{pq} and \gls{qapa} AoI \gls{pmf} over $t_q$, $\epsilon=0.7$.}
        \label{fig:matrix_07_te}
    \end{subfigure}	
    \begin{subfigure}[b]{.45\linewidth}
	    \centering
        \input{RevisionFigures/CDF,error=0.7,Tq=40,Pb=0.1,X=10,case2}  
        \caption{Complementary CDF, $\epsilon=0.7$.}
        \label{fig:CDF_0.7_40_01_te10}
    \end{subfigure}
 \caption{AoI distributions and CCDFs for \gls{pq} and \gls{qapa} for $T_q=40,\mu_b=0.1$ and $\epsilon=\{0.2,0.7\}$.}\vspace{-0.5cm}
 \label{fig:CDF_te}
\end{figure*}

It is possible to imagine a simple threshold-based strategy that performs almost as well as the optimal strategy in this scenario: sending packets only if $b\geq T_q-t_q$ is a good approximation of the optimal strategy. Computing the threshold \gls{aoi} over which it is convenient to send the first packet earlier is more complex, but still relatively easy in this case. More complex cases, such as the ones we will examine below, have a much wider state space with a higher dimensionality, making strategies difficult to visualize and hand-design.

\subsection{Periodic queries and error probability}

We now analyze what happens when the error probability is not constant, but follows a periodic function. We consider the case of a \gls{leo} satellite connection which has limited availability due to constraints on the available constellation: connectivity is only available sporadically, depending on the periodic passes of the satellite over the transmitter node. In order to show the main trade-offs and represent a realistic case in which a \gls{leo} satellite passes over the transmitter at regular intervals, we consider a periodic error process with period $T_e$, in which the state space $\mathcal{S}_e=\{1,\ldots,T_e\}$. The first two slots of each period are the only ones during which a transmission is possible, with an error probability $\epsilon_0$, and correspond to the satellite pass. In all other slots, the transmission fails, as the transmitter is outside the satellite's coverage area. We then have $\epsilon(1)=\epsilon(2)=\epsilon_0$, and $\epsilon(s_e)=1\,\forall s_e\notin\{1,2\}$. The transition probabilities for the error probability process are given by:
\begin{equation}
P_e(s_e,s_e')=\begin{cases}
               1, &s_e<T_e\wedge s_e'=s_e+1;\\
               1, &s_e=T_e\wedge s_e'=1;\\
               0, &\text{otherwise.}
              \end{cases}
\label{eq:det_error}
\end{equation}

 We can analyze the behavior of the system as a function of the period $T_e$ and the basic error probability $\epsilon_0$. The query arrival process is defined as above, with transition probabilities given by~\eqref{eq:det_queries}.
We first consider what happens in a simple case, setting $\mu_b=0.1$, $T_e=10$, and $T_q=40$. Fig.~\ref{fig:CDF_0.2_40_01_te10} and Fig.~\ref{fig:CDF_0.7_40_01_te10} show how the \gls{pq} and \gls{qapa} system are restricted to the available transmission slots: outside of those slots, the \gls{aoi} can only grow, resulting in the striped pattern on the plots. As expected, the \gls{qapa} system concentrates its effort in the transmission opportunities closer to a query, while the \gls{pq} system uses all available slots indiscriminately. This generates the difference in the \gls{qaoi} seen in Fig.~\ref{fig:CDF_0.2_40_01_te10} and Fig.~\ref{fig:CDF_0.7_40_01_te10}: the \gls{qapa} system can maintain a lower \gls{qaoi}, and the higher percentiles of the \gls{aoi}, i.e., the tail of its distribution, are lower as well if the error probability is high.

\begin{figure*}[t!]
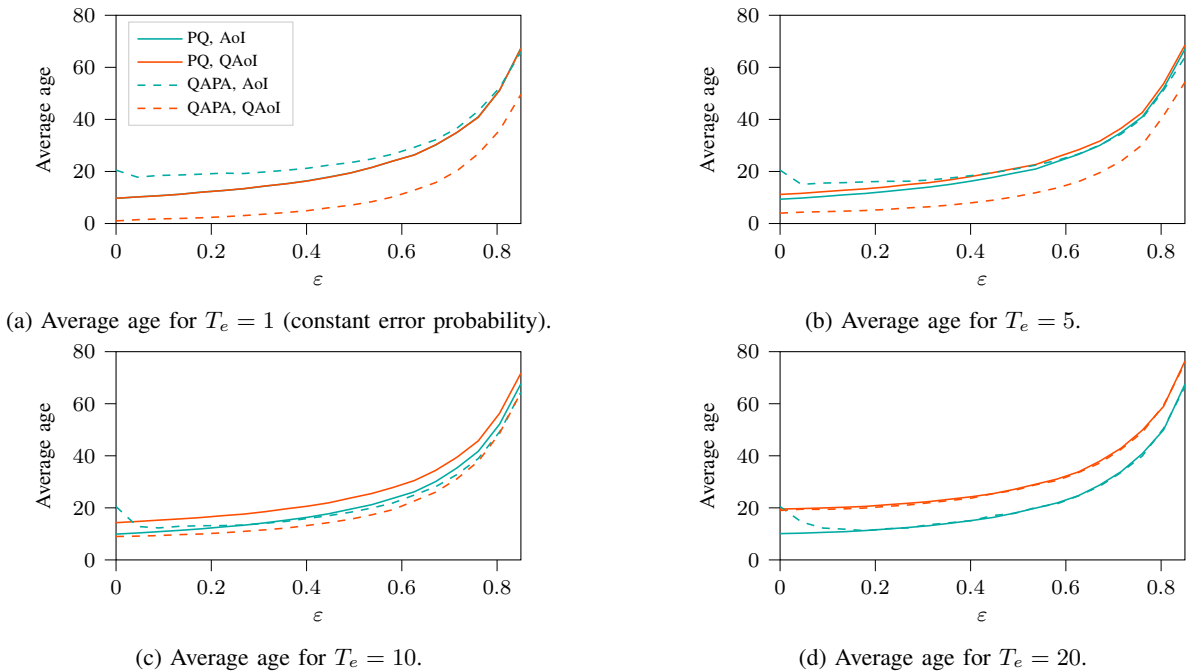

	\centering
	\begin{subfigure}[b]{0.48\linewidth}
	    \centering
        \input{RevisionFigures/AverageAge,Tq=40,Pb=0.1,X=1,case2}      
        \caption{Average age for $T_e=1$ (constant error probability).}
        \label{fig:avg_te1}
    \end{subfigure}	
    \begin{subfigure}[b]{0.48\linewidth}
	    \centering
        \input{RevisionFigures/AverageAge,Tq=40,Pb=0.1,X=5,case2}     
        \caption{Average age for $T_e=5$.}
        \label{fig:avg_te5}
    \end{subfigure}
	\begin{subfigure}[b]{0.48\linewidth}
	    \centering
        \input{RevisionFigures/AverageAge,Tq=40,Pb=0.1,X=10,case2}     
        \caption{Average age for $T_e=10$.}
        \label{fig:avg_te10}
	\end{subfigure}
		\begin{subfigure}[b]{0.48\linewidth}
	    \centering
        \input{RevisionFigures/AverageAge,Tq=40,Pb=0.1,X=20,case2}      
        \caption{Average age for $T_e=20$.}
        \label{fig:avg_te20}
    \end{subfigure}	
 \caption{Average \gls{aoi} and \gls{qaoi} for the two systems for different values of $T_e$, with $\mu_b=0.1$ and $T_q=40$.}\vspace{-0.5cm}
 \label{fig:average_age_te}
\end{figure*}

\begin{figure*}
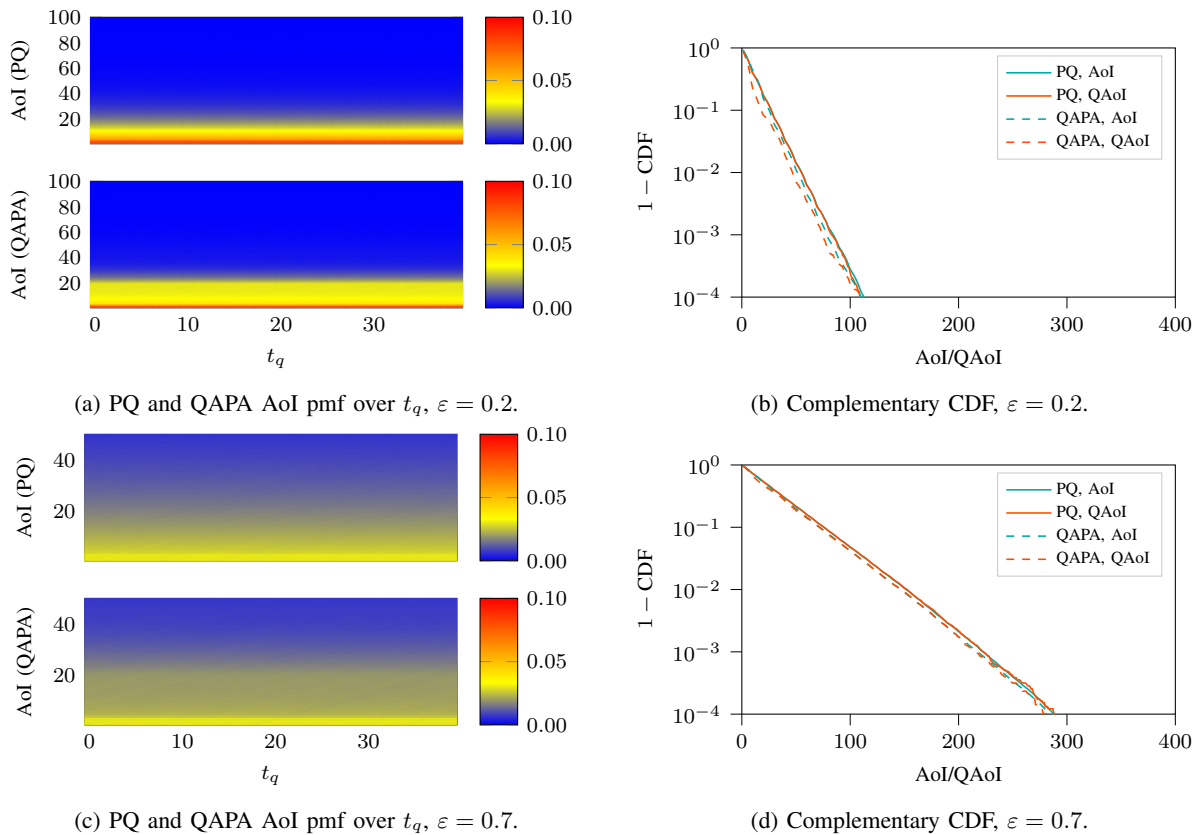

    \centering
	\begin{subfigure}[b]{.45\linewidth}
	    \centering
	    %\tikzexternalenable
        \input{RevisionFigures/matrix_02_unif20_case3}
        %\tikzexternaldisable
        \caption{\gls{pq} and \gls{qapa} AoI \gls{pmf} over $t_q$, $\epsilon=0.2$.}
        \label{fig:matrix_02_unif20}
    \end{subfigure}
	\centering
	\begin{subfigure}[b]{.45\linewidth}
	    \centering
        \input{RevisionFigures/CDF,error=0.2,Tq=40,Pb=0.1,X=10,case3}    
        \caption{Complementary CDF, $\epsilon=0.2$.}
        \label{fig:CDF_0.2_40_01_unif20}
    \end{subfigure}	
	\begin{subfigure}[b]{.45\linewidth}
	    \centering
	    %\tikzexternalenable
        \input{RevisionFigures/matrix_07_unif20_case3}
        %\tikzexternaldisable
        \caption{\gls{pq} and \gls{qapa} AoI \gls{pmf} over $t_q$, $\epsilon=0.7$.}
        \label{fig:matrix_07_unif20}
    \end{subfigure}	
    \begin{subfigure}[b]{.45\linewidth}
	    \centering
        \input{RevisionFigures/CDF,error=0.7,Tq=40,Pb=0.1,X=10,case3}     
        \caption{Complementary CDF, $\epsilon=0.7$.}
        \label{fig:CDF_0.7_40_01_unif20}
    \end{subfigure}
 \caption{AoI distributions and CCDFs for \gls{pq} and \gls{qapa} for $T_q=40,\mu_b=0.1,T_e=10$, and $\epsilon=\{0.2,0.7\}$, with uniformly distributed query intervals over $\{21,22,\ldots,40\}$.}\vspace{-0.5cm}
 \label{fig:CDF_unif20}
\end{figure*}

We can then look at the effect of increasing the period of the satellite on the interplay between \gls{aoi} and \gls{qaoi}. We note here that we consider the worst possible scenario for the \gls{qaoi}, i.e., the one in which the queries are synchronized with the satellite passes and each query arrives at the instant immediately before a satellite's pass. Fig.~\ref{fig:average_age_te} shows the difference in the average age for $T_e=1$ (i.e., the previous scenario with constant error probability), $T_e=5$, $T_e=10$, and $T_e=20$, considering $\mu_b=0.05$ and $T_q=40$. In all cases, the average \gls{aoi} of the \gls{pq} process is similar: since the most important limiting factor is the energy constraint, the average age is about 20 slots in the error-free case and follows a similar trend for all subfigures. By comparing Fig.~\ref{fig:avg_te1} and Fig.~\ref{fig:avg_te20}, it is clear that this is not true for \gls{qaoi}: the effect of having transmissions only at the beginning of the period, at least $T_e-2$ slots from the query, increases the average \gls{qaoi} for the \gls{pq} process by approximately $T_e/2-1$ slots. As in the previous case, the \gls{qapa} system can improve the \gls{qaoi} by paying a small cost in terms of \gls{aoi}, but the difference between its \gls{qaoi} and the \gls{pq} system's reduces as transmission opportunities become scarcer: by constraining the possible transmissions of the \gls{qapa} system to a few slots, we reduce the optimality gap of the traditional \gls{aoi} maximization strategy. However, the difference between the two is still significant even for $T_e=20$, as shown in Fig.~\ref{fig:avg_te20}. It is also interesting to see that \gls{aoi} and \gls{qaoi} are not the same in this case, even for the \gls{pq} policy: this is due to the fact that, as $T_q$ is a multiple of $T_e$, queries always come just before a transmission opportunity, so the \gls{qaoi} is always at least $T_e-1$, while there is no such offset for \gls{aoi}, which is measured in all slots.

\subsection{Stochastic queries with periodic error probability}

We now examine a more general case, in which queries arrive at stochastic \gls{iid} intervals with a known distribution and transmission opportunities are limited by satellite passes. The error probability process is then defined as above, with transition probabilities given by~\eqref{eq:det_error}. On the other hand, the queries are modeled to arrive with uniformly distributed inter-query times, i.e., $t_{q,i+1}-t_{q,i}\sim \mathcal{U}\left(\frac{T_q}{2}+1,T_q\right)$. This is a worst-case scenario for the \gls{qapa} system: as queries can arrive at a random instant over a wide range of values, the transmitter needs to keep the \gls{aoi} low almost at all times. We have $\mathcal{S}_q=\{1,\ldots,T_q\}$ with these non-zero transition probabilities:
\begin{equation}
  P_q(s_q,s_q')=\begin{cases}
1, &s_q\leq\frac{T_q}{2}\wedge s_q'=s_q+1;\\
1-\frac{1}{T_q-s_q+1}, &\frac{T_q}{2}<s_q<T_q\wedge s_q'=s_q+1;\\
\frac{1}{T_q-s_q+1}, & s_q>\frac{T_q}{2}\wedge s_q'=1.
\end{cases}    \label{eq:stochastic_q}
\end{equation}

\begin{figure*}[t!]
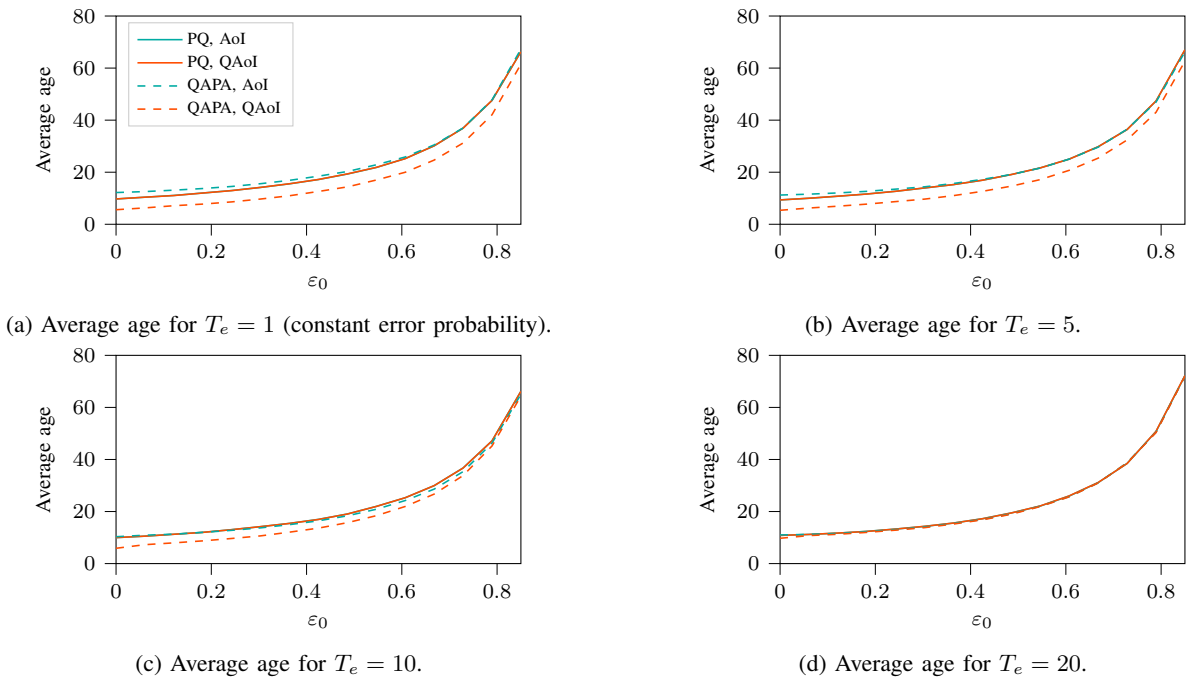

	\centering
	\begin{subfigure}[b]{0.48\linewidth}
	    \centering
        \input{RevisionFigures/AverageAge,Tq=40,Pb=0.1,X=1,case3}     
        \caption{Average age for $T_e=1$ (constant error probability).}
        \label{fig:avg_te1_unif20}
    \end{subfigure}	
    \begin{subfigure}[b]{0.48\linewidth}
	    \centering
        \input{RevisionFigures/AverageAge,Tq=40,Pb=0.1,X=5,case3}
        \caption{Average age for $T_e=5$.}
        \label{fig:avg_te5_unif20}
    \end{subfigure}
	\begin{subfigure}[b]{0.48\linewidth}
	    \centering
        \input{RevisionFigures/AverageAge,Tq=40,Pb=0.1,X=10,case3}      
        \caption{Average age for $T_e=10$.}
        \label{fig:avg_te10_unif20}
	\end{subfigure}
		\begin{subfigure}[b]{0.48\linewidth}
	    \centering
        \input{RevisionFigures/AverageAge,Tq=40,Pb=0.1,X=20,case3}      
        \caption{Average age for $T_e=20$.}
        \label{fig:avg_te20_unif20}
    \end{subfigure}	
 \caption{Average \gls{aoi} and \gls{qaoi} for the two systems for different values of $T_e$, with $\mu_b=0.1$, $T_q=40$, and uniformly distributed query intervals over $\{21,22,\ldots,40\}$.}\vspace{-0.5cm}
 \label{fig:average_age_unif20}
\end{figure*}

Fig.~\ref{fig:CDF_unif20} shows that this is indeed a case in which knowledge of the query process is not critical: the age \glspl{pmf} in Fig.~\ref{fig:matrix_02_unif20} and Fig.~\ref{fig:matrix_07_unif20} are almost the same for \gls{pq} and \gls{qapa}. In this case, the time since the last query has a limited value to the \gls{qapa} system, as it does not help much in predicting when a query will arrive. Consequently, the behavior of the \gls{qapa} system is much more similar to the \gls{pq} system's: the knowledge of the query arrival process statistics results in a very small gain, as the uniform interval distribution gives a limited amount of information on future steps. However, setting a uniform distribution for the interval still implies some memory in the process, since, e.g., knowing that there has been no query for $T_q-1$ steps implies that a query will arrive in the next step with probability 1. The \gls{pq} and \gls{qapa} strategies would be absolutely identical if the query process was Poisson, i.e., memoryless: if queries were to follow a memoryless process, any instant would be as valuable as any other in terms of future \gls{qaoi}. This is evident in Fig.~\ref{fig:CDF_0.2_40_01_unif20}, which shows a negligible gain for the \gls{qapa} system in terms of \gls{qaoi}, and even more in Fig.~\ref{fig:CDF_0.7_40_01_unif20}.

The analysis of the average \gls{aoi} and \gls{qaoi} as a function of the error probability $\epsilon_0$, shown in Fig.~\ref{fig:average_age_unif20}, shows that freedom of action and the precision of knowledge about the query arrival times are two factors that increase the gap between a naive \gls{pq} system and a query-aware \gls{qapa} one. This is intuitive, as limits to the available strategies can reduce gains, as can uncertainty about query arrival times. The more randomness is included in the system, and the more constrained the possible strategies become, the more \gls{qaoi} looks exactly like \gls{aoi}.

\subsection{Stochastic queries with a Gilbert-Elliott channel}

Finally, we consider another case, in which the channel does not have predefined transmission opportunities, but follows a stochastic Gilbert-Elliott~\cite{gilbert1960capacity,elliott1963estimates} model. In this model, we have two states, with $\varepsilon_1=0.2$ and $\varepsilon_2=0.7$. The transition probability matrix is given by:
\begin{equation}
 P_e(s_e,s_e')=\begin{pmatrix}
0.9 & 0.1\\
0.6 & 0.4
\end{pmatrix}.
\end{equation}
On the other hand, the query process is given by the distribution in~\eqref{eq:stochastic_q}. In this case, both the channel and the query process are stochastic, and the amount of information available to the sensor is limited to the current state of each process and the transition probabilities.

\begin{figure*}
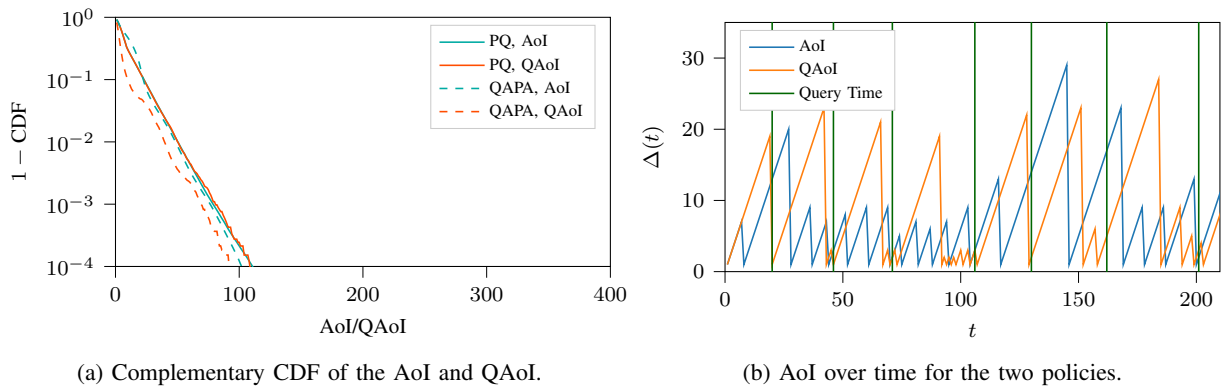

    \centering
    	\begin{subfigure}[t]{.45\linewidth}
	    \centering
    \input{RevisionFigures/GE_CDF}\vspace{-0.42cm}
        \caption{Complementary CDF of the \gls{aoi} and \gls{qaoi}.}
        \label{fig:GE_CDF}
    \end{subfigure}	
    	\begin{subfigure}[t]{.45\linewidth}
	    \centering
        \input{RevisionFigures/GE_Sawtooth}
        \caption{\gls{aoi} over time for the two policies.}
        \label{fig:GE_saw}
    \end{subfigure}	
    \caption{Behavior of the two systems with stochastic queries and a Gilbert-Elliott channel.}\vspace{-0.5cm}
    \label{fig:gilbert}
\end{figure*}

Fig.~\ref{fig:GE_CDF} shows the complementary \gls{cdf} of the \gls{aoi} and \gls{qaoi} in this scenario: as this case is less restrictive than the one with periodic transmissions, in which a successful update was only possible once every 5 or more slots, the \gls{qapa} strategy performs significantly better in terms of \gls{qaoi}. Surprisingly, it also performs better in terms of the tail of the \gls{aoi}, although not on average: this might be because of the choice to save energy by not transmitting in the first 20 slots after each query, which have a probability 0 of having another query. Fig.~\ref{fig:GE_saw} shows this behavior over time: while the \gls{pq} policy tends to uniformly transmit frequently, the \gls{qapa} policy lets the \gls{aoi} grow when queries cannot arrive, then transmits more often to maintain the lowest possible \gls{qaoi} for a potential query.

\section{Conclusions and future work}\label{sec:concl}

In this work, we have presented an optimization of \gls{qaoi}, which takes the query arrival process and resource constraints on the communication into account. We showed that awareness of the query process can improve average and worst-case freshness in a variety of systems, modeling the single-source scheduling problem as an \gls{mdp} and finding the analytical solution. As \gls{aoi} does not consider the specific features of applications, but reduces the age of any packet at any time, it cannot incorporate this additional information.  The awareness of the query process can significantly improve the freshness as perceived by several monitoring application, adapting the scheduling to only transmit when it is most useful and avoid useless updates.

This work is a first step in considering the requirements of time-sensitive monitoring applications in resource-constrained communication scenarios: we see several avenues of possible future work, such as including the value of updates in the scheduling problem as well as their timing. Furthermore, the extension of the problem to more complex systems with multiple sources and realistic channel access can be an interesting direction of research, as there are several scenarios with one or more monitoring applications that need information from multiple sensors. In these more complex scenarios, policy iteration would be too complex due to the curse of dimensionality, and we foresee the application of reinforcement learning methods to find the optimal strategy for minimizing \gls{qaoi}.

\appendix

In this Appendix, we prove that the threshold strategies defined in Sec.~\ref{ssec:threshold} are optimal. In order to prove Theorem~\ref{th:threshold}, we first introduce the expected long-term cost $\Gamma(t,a,d)$:
\begin{equation}
 \Gamma(t,a,d)=\mathbb{E}[C(t)|a_{t}=a,\Delta(t-1)=d].
\end{equation}
We can similarly define $\Gamma(t,\pi,d)$:
\begin{equation}
 \Gamma(t,\pi,d)=\mathbb{E}[C(t)|\pi,\Delta(t-1)=d].
\end{equation}
We then introduce the following lemma.
\begin{lemma}\label{th:monot}
The expected long-term cost is a monotonically increasing function of the age $\Delta$:
\begin{equation}
  \Gamma(t,a,d)\leq\Gamma(t,a,d+1),\ \forall a,d,t.
\end{equation}
\end{lemma}
\begin{proof}
 We prove this lemma by backward induction in $t$, working from $t=T$ and going backward. If we consider the base case with $t=T$, we can see from~\eqref{eq:lastreward} that the condition holds. We now assume that the condition is from $t+1$ to $t=T$. First, we consider the case with $a_{t}=0$. We have:
 \begin{equation}
 \Gamma_{\text{PQ}}(t,0,d)=d+\Gamma_{\text{PQ}}(t+1,\pi^*,d+1).
 \end{equation}
Since we know that $\Gamma_{\text{PQ}}(t+1,\pi^*,d+2)\geq\Gamma_{\text{PQ}}(t+1,\pi^*,d+1)$ due to the assumption in the inductive step, and $d+1>d$, the lemma is proven for $a_{t}=0$. If we consider the case in which $a_{t}=1$, we get:
\begin{equation}
\begin{aligned}
     \Gamma_{\text{PQ}}(t,1,d)=&1+(1-\varepsilon)\Gamma_{\text{PQ}}(t+1,\pi^*,1)\\
     &+\varepsilon (d+\Gamma_{\text{PQ}}(t+1,\pi^*,d+1)).
\end{aligned}
\end{equation}
It is easy to see how this case also respects the condition, as $d$ appears twice. All elements of the cost are then the same or higher for $\Delta(t-1)=d+1$ and the lemma is proven by induction. The same procedure can be repeated for \gls{qapa}, using either \gls{qaoi} or \gls{eaoi} as a metric. 
\end{proof}
As the expected cost is a monotonically increasing function of the current age, we can immediately prove Theorem~\ref{th:threshold}.
\renewcommand*{\proofname}{Proof of Theorem~\ref{th:threshold}}
\begin{proof}
As for Lemma~\ref{th:monot}, we can use backward induction starting from $t=T$ to prove the theorem. The base case is trivially true, as the decision is given by~\eqref{eq:lastslot}. We can now attempt to perform the inductive step \emph{ad absurdum}, by assuming that the theorem is true from $t+1$ to $T$. Suppose that the optimal policy is not a threshold policy, i.e., $\exists d:\pi^*(d,t)=1\wedge\pi^*(d+1,t)=0$. In this case, we know that, since $\pi^*(d,t)=1$, we get:
\begin{equation}
\Gamma_{\text{PQ}}(t,1,d)<\Gamma_{\text{PQ}}(t,0,d).
\end{equation}
We can then take the components of the reward:
\begin{equation}
\begin{aligned}
     1+d+\Gamma_{\text{PQ}}(t+1,\pi^*,d+1)>(1-\varepsilon)\Gamma_{\text{PQ}}(t+1,\pi^*,1)\\
     +1+\gamma+\varepsilon (d+\Gamma_{\text{PQ}}(t+1,\pi^*,d+1)),
\end{aligned}
\end{equation}
which yields:
\begin{equation}
      d+\Gamma_{\text{PQ}}(t+1,\pi^*,d+1)-\Gamma_{\text{PQ}}(t+1,\pi^*,1)>\frac{\gamma}{1-\varepsilon}.\label{eq:txd}
\end{equation}
On the other hand, since we assumed that $\pi^*(d+1,t)=0$, we can follow the same procedure to get:
\begin{equation}
        d+1+\Gamma_{\text{PQ}}(t+1,\pi^*,d+2)-\Gamma_{\text{PQ}}(t+1,\pi^*,1)\leq\frac{\gamma}{1-\varepsilon}.\label{eq:notxd}
\end{equation}
Since $C(t)$ is non-negative, and we know from Lemma~\ref{th:monot} that $\Gamma_{\text{PQ}}(t+1,\pi^*,d+1)\leq\Gamma_{\text{PQ}}(t+1,\pi^*,d+2)$, the results in~\eqref{eq:txd} and~\eqref{eq:notxd} are in contradiction, and $\pi^*(d,t)=1\Rightarrow\pi^*(d+n,t)=1,\,\forall n\in\mathbb{N}$. As before, the same procedure can be repeated for \gls{qapa}, which does not have the term $d$ in the long-term cost, but still yields the same inequality.
\end{proof}

\bibliographystyle{IEEEtran}
\bibliography{journal_final}

\begin{IEEEbiography}[{\includegraphics[width=1in,height=1.25in,clip,keepaspectratio]{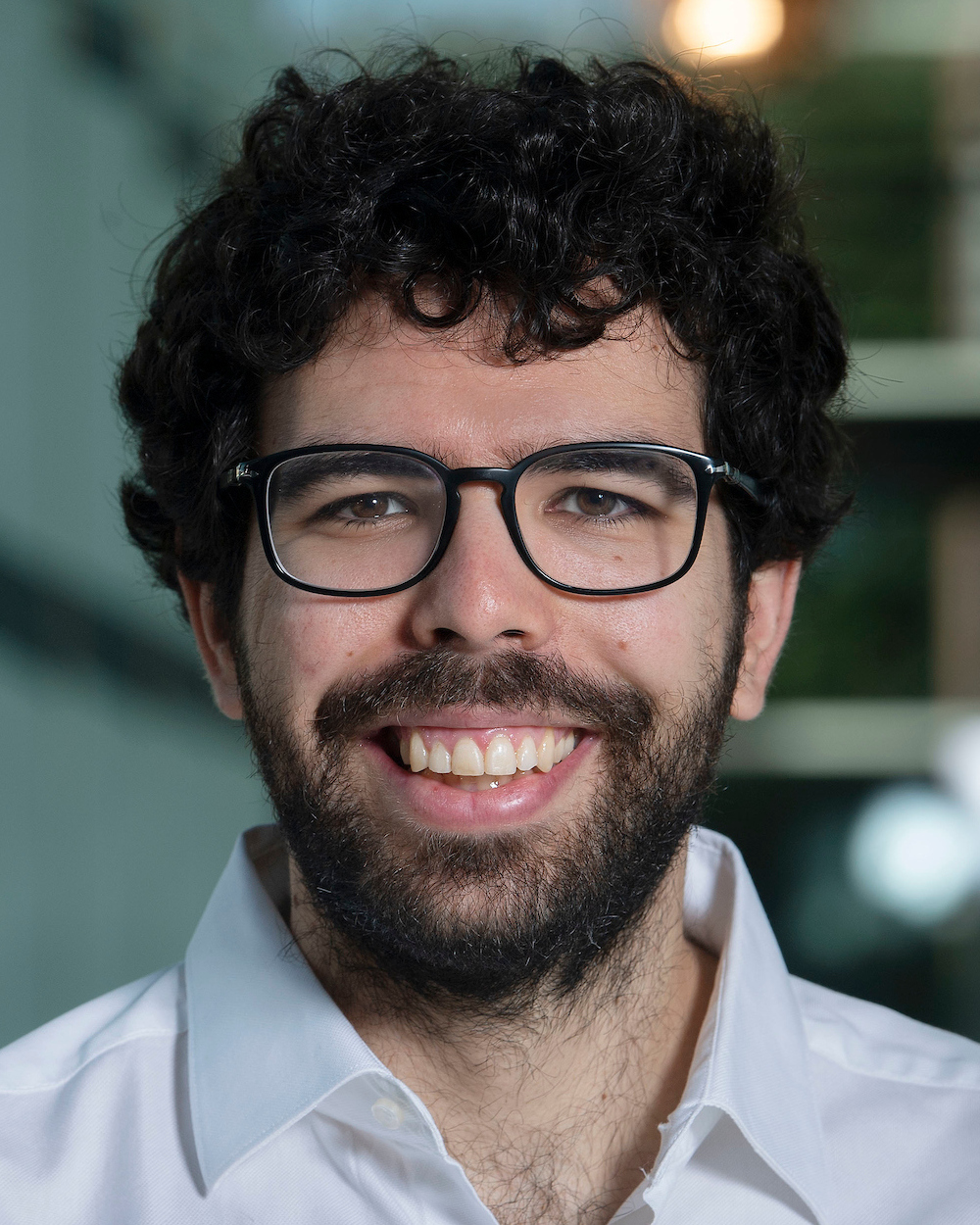}}]
{Federico Chiariotti} (S'15--M'19) is currently a post-doctoral researcher at the Department of Electronic Systems, Aalborg University, Denmark. He received his Ph.D. in information engineering in 2019 from the University of Padova, Italy. He received the bachelor's and master's degrees in telecommunication engineering (both \emph{cum laude}) from the University of Padova, in 2013 and 2015, respectively. He has authored over 40 published papers on wireless networks and the use of artificial intelligence techniques to improve their performance. He was a recipient of the Best Paper Award at several conferences, including the IEEE INFOCOM 2020 WCNEE Workshop. His current research interests include network applications of machine learning, transport layer protocols, Smart Cities, bike sharing system optimization, and adaptive video streaming.
\end{IEEEbiography}

\begin{IEEEbiography}[{\includegraphics[width=1in,height=1.25in,clip,keepaspectratio]{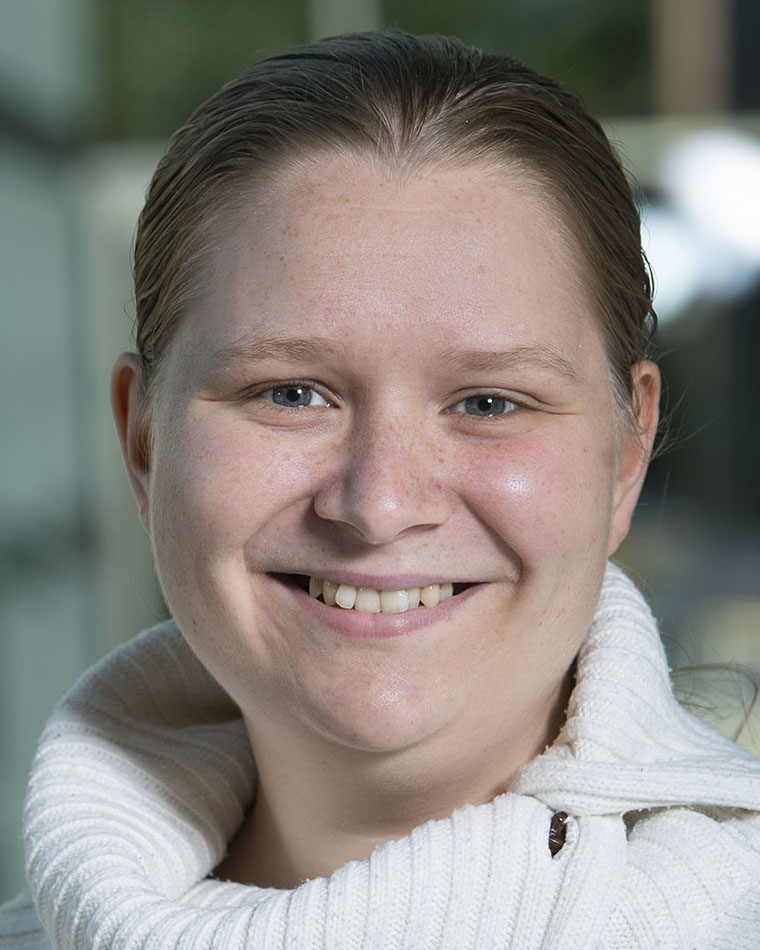}}]
{Josefine Holm} received her B.Sc and M.Sc. degrees in mathematical engineering from Aalborg University in 2016 and 2018, respectively. She is currently pursuing her Ph.D. degree at the Connectivity Section at Aalborg University. Her research interests include wireless communication and IoT networks.
\end{IEEEbiography}

\begin{IEEEbiography}[{\includegraphics[width=1in,height=1.25in,clip,keepaspectratio]{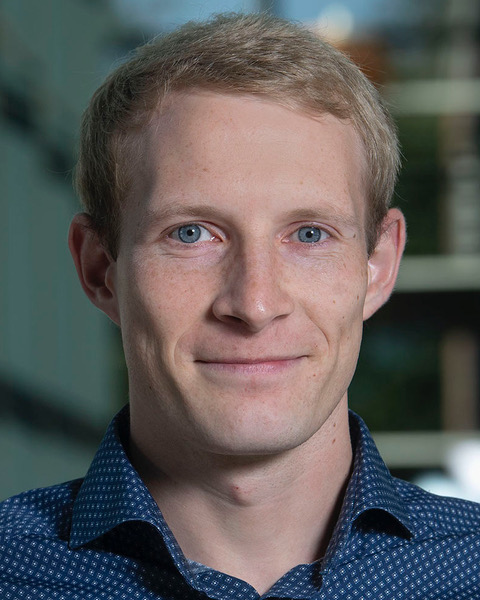}}]
{Anders E. Kal\o r}
(S'17) received the B.Sc. degree in computer engineering and the M.Sc. degree in networks and distributed systems from Aalborg University, Denmark, in 2015 and 2017, respectively. He is currently pursuing a Ph.D. degree in the area of wireless communications and networking at the Connectivity section at Aalborg University. In 2017, he was a visiting student at Robert Bosch, Germany, and in 2020 at King's College London, UK. His research interests include communication theory, MAC layer design for wireless systems, and networking.
\end{IEEEbiography}

\begin{IEEEbiography}[{\includegraphics[width=1in,height=1.25in,clip,keepaspectratio]{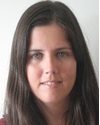}}]
{Beatriz Soret}
(M'11) received her M.Sc. and Ph.D. degrees in Telecommunications from the University of Malaga, Spain, in 2002 and 2010, respectively. She is currently an associate professor at the Department of Electronic Systems, Aalborg University, and a Senior Research Fellow at the Communications Engineering Department, University of Malaga. Her current research interests include semantic communications and AoI, LEO satellite communications, and intelligent IoT environments.
\end{IEEEbiography}

\begin{IEEEbiography}[{\includegraphics[width=1in,height=1.25in,clip,keepaspectratio]{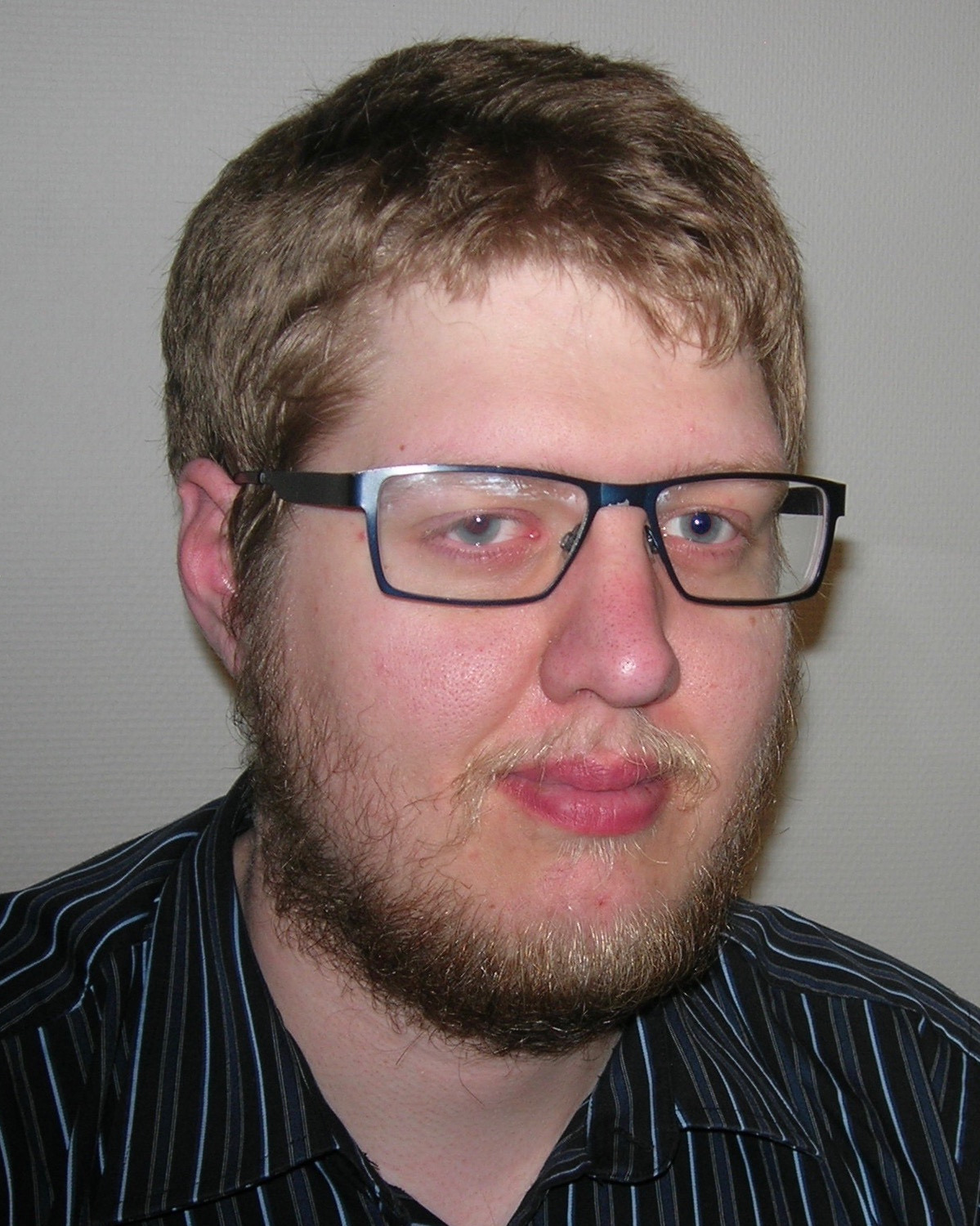}}]
{S\o ren K. Jensen}
is a postdoctoral researcher at the Center for Data-Intensive Systems (Daisy) at the Department of Computer Science, Aalborg University, Denmark. His research interests span multiple areas of Computer Science, including programming language design, compiler design and implementation, developer tooling, parallel and distributed computing, big data, database management systems, data warehousing, and extract-transform-load processes.
\end{IEEEbiography}

\begin{IEEEbiography}[{\includegraphics[width=1in,height=1.25in,clip,keepaspectratio]{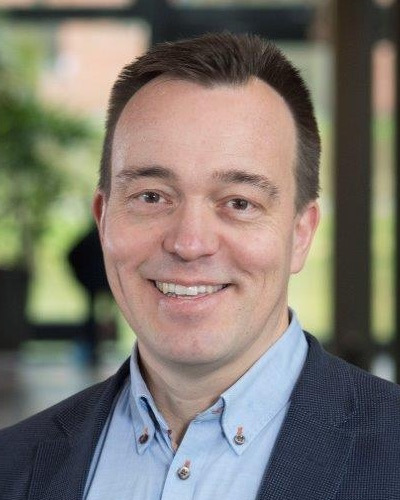}}]
{Torben B. Pedersen} is a professor
with the Center for Data-Intensive Systems
(Daisy), Aalborg University, Denmark. His
research concerns Predictive, Prescriptive, and Extreme-Scale Data Analytics with Digital Energy as the main application area.
He is an ACM Distinguished Scientist, a Senior Member of the
IEEE, a member of the Danish Academy of
Technical Sciences, and holds an honorary doctorate from TU Dresden. 
\end{IEEEbiography}

\begin{IEEEbiography}[{\includegraphics[width=1in,height=1.25in,clip,keepaspectratio]{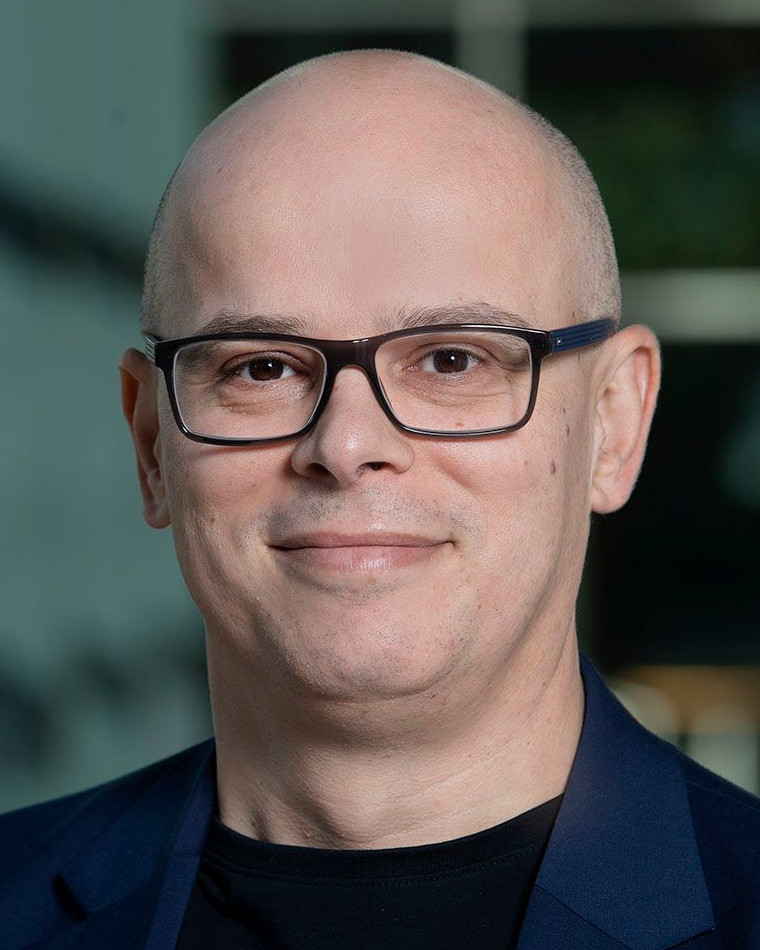}}]
{Petar Popovski} (S'97--A'98--M'04--SM'10--F'16)
is a Professor at Aalborg University, where he heads the section on Connectivity and a Visiting Excellence Chair at the University of Bremen. He received his Dipl.-Ing and M. Sc. degrees in communication engineering from the University of Sts. Cyril and Methodius in Skopje and the Ph.D. degree from Aalborg University in 2005. He is a Fellow of the IEEE. He received an ERC Consolidator Grant (2015), the Danish Elite Researcher award (2016), IEEE Fred W. Ellersick prize (2016), IEEE Stephen O. Rice prize (2018), Technical Achievement Award from the IEEE Technical Committee on Smart Grid Communications (2019), the Danish Telecommunication Prize (2020) and Villum Investigator Grant (2021). He is a Member at Large at the Board of Governors in IEEE Communication Society, Vice-Chair of the IEEE Communication Theory Technical Committee and IEEE TRANSACTIONS ON GREEN COMMUNICATIONS AND NETWORKING. He is currently an Editor-in-Chief of IEEEE JOURNAL ON SELECTED AREAS IN COMMUNICATIONS. Prof. Popovski was the General Chair for IEEE SmartGridComm 2018 and IEEE Communication Theory Workshop 2019. His research interests are in the area of wireless communication and communication theory. He authored the book ``Wireless Connectivity: An Intuitive and Fundamental Guide'', published by Wiley in 2020.
\end{IEEEbiography}

\end{document}